\documentclass[journal]{IEEEtran}

\usepackage{amsmath,amsfonts}
\usepackage[caption=false,font=footnotesize]{subfig}
\usepackage{textcomp}
\usepackage{stfloats}
\usepackage{url}
\usepackage{verbatim}
\usepackage{graphicx}
\usepackage{cite}
\usepackage{setspace}
\usepackage{epsfig}  
\usepackage{amssymb}  
\usepackage{lscape}  
\usepackage{color}
\usepackage{mathrsfs}
\usepackage{mathtools}
\usepackage{amsthm}
\usepackage{array}
\usepackage{booktabs}
\usepackage{multicol}
\usepackage{gensymb}
 
\usepackage{longtable}
 
\usepackage{cases}
\usepackage{bm}
\usepackage{textcomp}
\usepackage{upgreek}

\usepackage{array}

\begin{document}
%
\title{Singular Vector Finite Element Basis Functions for Tetrahedra in Complex Electromagnetic Geometries}
%
%
%

\author{Samuel T. Elkin,~\IEEEmembership{Graduate Student Member,~IEEE,}, Ghazi Khan,~\IEEEmembership{Graduate Student Member,~IEEE,} Ebrahim Forati,~\IEEEmembership{Senior Member,~IEEE,} Brandon W. Langley, Dogan Timucin, Reza Molavi,~\IEEEmembership{Member,~IEEE,}
and~Thomas E. Roth,~\IEEEmembership{Senior Member,~IEEE}
\thanks{Manuscript received XX XXX XXXX; revised XX XXX XXXX; accepted XX XXX XXXX. This work was supported by a gift from Google LLC. (\textit{Corresponding author: Thomas E. Roth})\\
\indent Samuel T. Elkin, Ghazi Khan, and Thomas E. Roth are with the Elmore Family School of Electrical and Computer Engineering, Purdue University, West Lafayette, IN 47907 USA and the Purdue Quantum Science and Engineering Institute, West Lafayette, IN 47907 USA (e-mail: rothte@purdue.edu).

\indent Ebrahim Forati, Brandon Langley, Dogan Timucin, and Reza Molavi are with Google Quantum AI, Santa Barbara, California 93111, USA}.}

%
%

\markboth{IEEE Transactions on Microwave Theory and Techniques}%
{Shell \MakeLowercase{\textit{et al.}}: Bare Demo of IEEEtran.cls for IEEE Journals}
%



\maketitle

\begin{abstract}
Electromagnetic finite element method (FEM) implementations using traditional vector basis functions typically struggle to accurately represent field behavior near singular features such as conducting wedges. To combat this, previous works have introduced specialized singular basis functions to directly model the singular fields in these regions, leading to substantially improved performance. While these efforts have been pursued extensively in 2D, relatively few functions have been developed for 3D elements. In this work, we develop a set of basis functions for modeling singular behavior in tetrahedra. Unlike prior functions for tetrahedra, these basis functions are additive, meaning they are included alongside the standard vector basis functions to achieve more robust performance. Further, these functions are designed to be adaptable to tetrahedra touching several unique singular features by using combinations of basis functions singular with respect to each node and edge in the element, making them applicable to highly complex geometries. Higher-order interpolatory versions of the basis functions for modeling singular behavior with greater accuracy are also provided. These basis functions are demonstrated to lead to substantial improvements in accuracy relative to the standard basis functions, and allow otherwise expensive simulations to be performed at far lower costs. As an application example, we perform simulations to extract critical quantities for designing superconducting quantum bits (qubits) that significantly depend on the behavior of singular fields. In Ansys HFSS, this took 21.27 hours and a peak memory usage of 6.23 TB with 800 processors available, while using our singular basis functions achieved comparable results in 196 seconds while using 27.24 GB of memory and only 16 processors. Due to these benefits, the singular basis functions developed here could be applied to enable design optimization of electromagnetic geometries with dominantly singular behavior, such as superconducting qubits.
\end{abstract}


%
\IEEEpeerreviewmaketitle

\section{Introduction}
\label{sec:intro}
\IEEEPARstart{S}{ingular} electromagnetic fields produced by sharp wedges have been studied for many decades. Early efforts in this endeavor focused on developing analytical solutions to describe the behavior around different types of singular features \cite{1972_Meixner_Singular, 1976_Hurd_Singular, 1985_Van-Bladel_Singular, 1996_VanBladel_Singular, 1988_Braver_Singular_Half-plane}. As computational methods such as the finite element method (FEM) and the method of moments (MoM) began to gain popularity, researchers found that standard basis functions tend to resolve these singular fields very poorly \cite{1986_Pantic_Singular_TX, 1977_Wilton_1D_Singular_MoM}. To improve this behavior, many researchers investigated the use of specialized basis functions for modeling singular fields, commonly referred to as singular basis functions. Many early contributions in this effort came from the mechanical engineering community \cite{1970_Byskov_Singular_Stress, 1971_Tracey_Singular_Stress, 1977_Tracey_Singular_Scalar, 1979_Stern_Singular_Scalar}, with some notable works from the electromagnetics community as well \cite{1986_Pantic_Singular_TX, 1994_Gil_Singular_Scalar}. 

Around the same time, N\'ed\'elec demonstrated the benefits of curl-conforming vector basis functions for electromagnetic finite element solutions \cite{1980_Nedelec_Vector_Basis}. To ensure compatibility with these functions, most subsequent efforts in developing singular basis functions for electromagnetic applications used vector functions. Many such functions were developed for triangles, with a mix of ``substitutive" functions which replaced the curl-conforming set in elements near the singularity \cite{1998_Pantic_Singular_Triangles, 2005_Sun_Singular_Tris, 2005_Masoni_Singular_Sub}, as well as some ``additive" functions that are included alongside N\'ed\'elec's functions \cite{1997_Gil_Singular_Triangles, 2004_Graglia_singular, 2014_Graglia_Singular_Hierarchical, 2014_Graglia_Corner_Singular}. Singular vector functions have also been explored for MoM with substitutive \cite{1999_Wilton_Singular_MoM, 2009_Ozturk_Singular_MoM} and additive \cite{2008_Graglia_Singular_MoM} functions in triangles, and more recent efforts in developing additive functions for quadrilateral elements \cite{2018_Graglia_Singular_Quadrilaterals, 2020_Graglia_Singular_Quadrilateral_Tip}. While both substitutive and additive functions can perform well in ideal circumstances, the additive versions tend to be more robust because they can rely on the curl-conforming functions when the singular functions alone would perform poorly. In contrast to the wealth of 2D basis functions, efforts in 3D elements such as tetrahedra have been sparse, and only substitutive functions have been explored to date \cite{2007_Webb_Singular_Tetrahedra, 2012_Webb_Singular_Hierarchical}. A more complete review of the topic can be found in \cite{2016_Peterson_Singular_Review}.

More recently, a potential application for these functions has emerged in the design of superconducting quantum computers \cite{2025_Roth_CEM_Review}. In this field, minimizing energy loss is of critical importance to designing quantum bits (qubits) with high coherence times \cite{krantz2019quantum}. Qubits are typically fabricated on substrates that exhibit incredibly small loss tangents at cryogenic temperatures, with modern processes achieving values as low as $\tan \delta = \mathcal{O}(10^{-8})$ \cite{2023_Read_Sapphire_Loss, 2024_Ganjam_ms_Coherence}. Conductive losses in the bulk regions of the superconductors are also minimal. Correspondingly, various studies have determined that the leading source of losses occur in the vicinity of $\mathcal{O}(1 \, \mathrm{nm})$ interfaces between bulk materials \cite{2014_Quintana_Interface_Losses, 2017_Gambetta_Surface_Loss, 2018_Caulsine_CPW_Interfaces, 2019_Woods_Interface_Loss, 2024_Ganjam_ms_Coherence}. As a result, accurately estimating the energy lost in these layers at the design stage is critical for fabricating high-performing qubits, and minimizing those losses is necessary for achieving high coherence times.

In the last decade, significant efforts have been devoted to better understand how to characterize these losses through simulations. The most common strategy is to use commercial FEM tools to extract a ``participation ratio," which describes the fraction of the total electrical energy in the geometry found within a given interface layer \cite{2011_Martinis_Participation}. If the material properties of the layer are known, they can be combined with the participation ratio to estimate the impact of each layer on qubit coherence times, informing the qubit design process. 

The primary challenge in this workflow is that commercial tools tend to produce incorrect participation ratios because they fail to properly resolve the contributions from singularities. In principle, this can be addressed with extremely fine meshing, but this comes at the cost of long simulations that make design optimization challenging. Many current efforts involve using 2D simulations to approximate the contribution of singularities, with these results then combined into those from a full 3D simulation with a much coarser mesh to estimate the full participation ratios \cite{2015_Devoret_Surface_Participation, 2024_Rodionov_Surface_Loss}. However, the process of combining the 2D and 3D results is inherently prone to error, making this approach nonideal. More recently, a surface integral equation solver for multilayer media has been developed to combat this issue \cite{2026_Wang_SIE}. While this tool can accurately evaluate participation ratios much more rapidly than with densely-meshed FEM, it cannot account for geometric complexities like substrate etching which are commonly used in this field to improve qubit coherence.

If specialized basis functions are used to model the singular fields, full-wave FEM simulations with a coarse mesh could be used to accurately extract participation ratios, circumventing these issues. While such functions have been proposed previously for tetrahedra \cite{2007_Webb_Singular_Tetrahedra, 2012_Webb_Singular_Hierarchical}, they are only applicable to tetrahedra with a single node or edge along a singular feature. Therefore, these functions are not suitable for the geometric complexity found in realistic devices, including for superconducting qubits. To address this issue, here we develop a new set of functions which are compatible with arbitrary configurations of tetrahedra. These are constructed by adapting functions developed for triangles \cite{2004_Graglia_singular} to tetrahedra, as well as introducing new functions for modeling the more complex singular behavior experienced in 3D simulations. Because superconducting qubit geometries typically only support singularities in the electric field, the functions we introduce here are designed to model a singular electric field and the associated nonsingular magnetic field. Unlike the prior functions for tetrahedra \cite{2007_Webb_Singular_Tetrahedra, 2012_Webb_Singular_Hierarchical}, these functions form an additive set to provide robust performance.

The remainder of the paper is structured as follows. In Section \ref{sec:formulation}, we first discuss the underlying theory behind the form of the singular functions. We then introduce functions capable of providing this behavior, and explain how to implement higher-order versions of them to improve accuracy. Various supporting details for these discussions are included in Appendices \ref{app:edge-sing} and \ref{app:edge-sing-rot}. In Section \ref{sec:results}, we validate the method against analytical solutions and examine behavior near singularities by comparing to results with standard FEM basis functions using extremely fine meshing in these regions. Additionally, we compare participation ratio calculations on a qubit geometry to Ansys HFSS to demonstrate the efficiency improvements enabled by these functions. We find that the HFSS simulation involves a memory-intensive $21.27 \, \mathrm{h}$ calculation, while using our singular basis functions allows us to achieve comparable results in only $196 \, \mathrm{s}$ with a low enough memory overhead to be easily run on a workstation. Accounting for the disparity in number of processors used between the two methods, an over 19,500-fold improvement in CPU-hours was achieved for this application.  Finally, the conclusions drawn from these results and areas for future work are discussed in Section \ref{sec:conclusion}.

\section{Formulation}
\label{sec:formulation}
In this section, we first describe the singular fields produced by a conducting wedge in Section \ref{subsec:singular}. This information is used in Section \ref{subsec:1st-order} to inform the construction of functions for modeling singular behavior emerging from nodes and edges of tetrahedra touching geometric features producing singular fields. These functions are additive, meaning they are included alongside standard vector basis functions. The sets of functions can be repeated for each singular feature within the tetrahedra, making them adaptable to highly complex 3D geometries. Procedures for generating higher-order functions of each type are also described in detail in Section \ref{subsec:higher-order}. We conclude in Section \ref{subsec:implementation} with a discussion of practical considerations for the efficient implementation of these functions.

\subsection{Singular Fields}
\label{subsec:singular}

\begin{figure}
    \centering
    \includegraphics[width=0.3\linewidth]{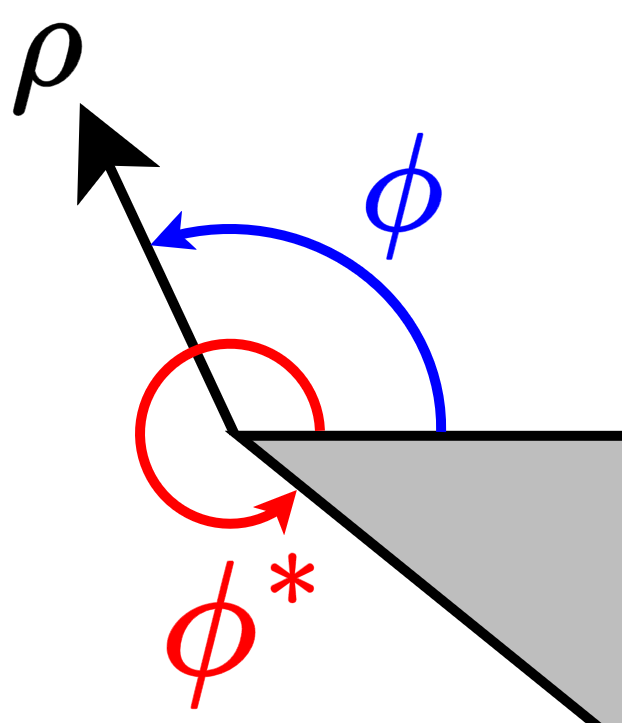}
    \caption{Cylindrical coordinate system defined around a sharp conductive wedge (colored gray) at angle $\phi^*$. The $z$-axis extends out of the page.}
    \label{fig:singEdge}
\end{figure}

The behavior of electromagnetic fields near sharp conducting edges is a topic which has received detailed analytical treatment \cite{1972_Meixner_Singular, 1976_Hurd_Singular, 1985_Van-Bladel_Singular, 1996_VanBladel_Singular, 1988_Braver_Singular_Half-plane}. The analysis is typically performed in a cylindrical coordinate system such as that of Fig. \ref{fig:singEdge}, where $\rho = 0$ is the position of the tip of a conducting wedge extending along the $z$-axis. Subsequently, we refer to the tip of this wedge as a ``singular feature" due to the singular fields produced by it. The fields surrounding the wedge can be described using the Meixner expansion \cite{1972_Meixner_Singular},
\begin{align}
    \mathbf{E}= \sum_{i=0}^\infty (a_i \boldsymbol{\hat{\rho}} + b_i \boldsymbol{\hat{\phi}} + c_i \mathbf{\hat{z}}) \rho^{\nu + i - 1}, 
    \label{eq:sing-E} \\
    \mathbf{H} = \sum_{i=0}^\infty (\alpha_i \boldsymbol{\hat{\rho}} + \beta_i \boldsymbol{\hat{\phi}} + \gamma_i \mathbf{\hat{z}}) \rho^{\nu + i - 1}.
    \label{eq:sing-H}
\end{align}
When the wedge is at an angle $\phi^*$ in a homogeneous region, $\nu = \pi/\phi^*$. If $\phi^* > \pi$, $1/2 \leq \nu < 1$ and the first term in the Meixner expansion becomes singular.

The expansions can be simplified by isolating modes with singular electric ($\alpha_0 = \beta_0 = 0$) or magnetic ($a_0 = b_0 = 0$) fields \cite{1976_Hurd_Singular, 1985_Van-Bladel_Singular}, where we focus only on modeling singular electric fields in this work as these are what typically occur for our application focus of superconducting qubits. An electrostatic analysis can be used to approximate the fields in the vicinity of the edge as \cite{1985_Van-Bladel_Singular, jackson1998classical}
\begin{align}
    \mathbf{E} \approx \nabla \Phi = (a_0&(\phi) \boldsymbol{\hat{\rho}} + b_0(\phi) \boldsymbol{\hat{\phi}}) \rho^{\nu-1},
    \label{eq:sing-E-approx} \\
    \mathbf{H} \approx \gamma_1 (\phi) \rho^\nu \mathbf{\hat{z}},
    \label{eq:sing-H-approx}
\end{align}
where $a_0$, $b_0$, and $\gamma_1$ depend on the neighboring geometry.

The expressions of (\ref{eq:sing-E-approx}) and (\ref{eq:sing-H-approx}) inform how we can construct basis functions to model singular behavior. In particular, we require a set of functions that can represent a ``gradient-singular" field with $\rho^{\nu-1}$ dependence with respect to the wedge. Additionally, we need ``rotational'' basis functions with $\rho^\nu$ dependence in their curl to properly model the magnetic field of (\ref{eq:sing-H-approx}). Then successive terms in the Meixner expansions of (\ref{eq:sing-E}) and (\ref{eq:sing-H}) can be represented using higher-order versions of those functions, which will be discussed in Section \ref{subsec:higher-order}.

We note that the field behavior described in this section is only valid for sharp wedges, and it does not reflect the field structure near 3D corners \cite{jackson1998classical}. In this work, we approximate the corner behavior with functions designed to model the fields in (\ref{eq:sing-E}) to (\ref{eq:sing-H-approx}), but specialized functions for modeling corner singularities could be developed in the future.

\subsection{First-Order Basis Functions}
\label{subsec:1st-order}

\subsubsection{Node-Singular Functions}
\label{subsubsec:node-funcs}

\begin{figure}[t!]
    \centering
    \includegraphics[width=0.7\linewidth]{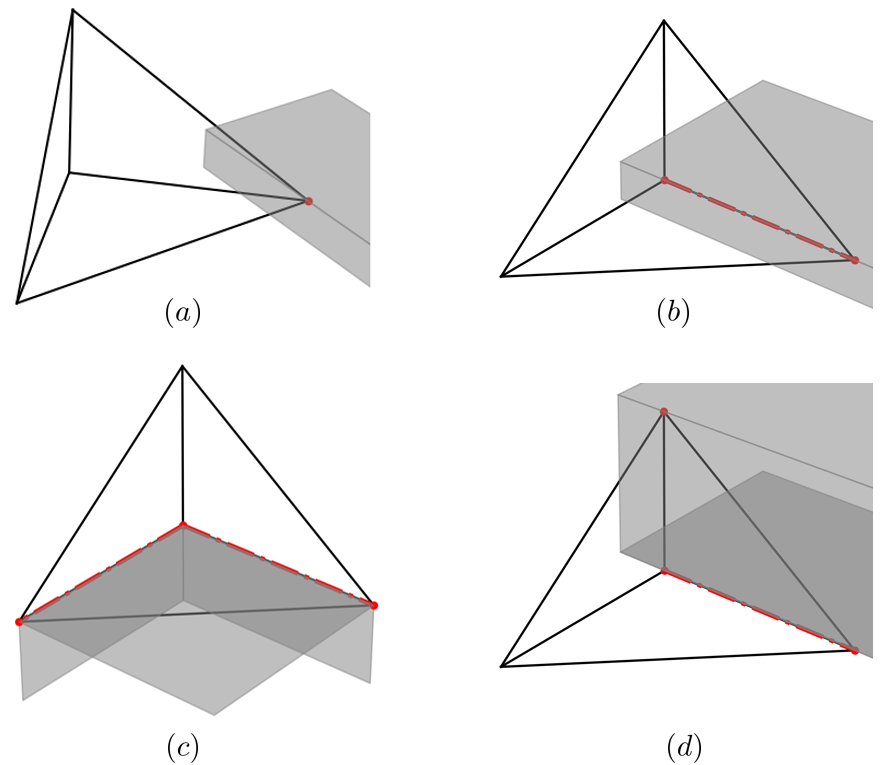}
    \caption{Examples of tetrahedra along singular features in a 3D geometry, with singular nodes and edges highlighted in red. In (a), the tetrahedron has only a single node along the conductive wedge, so only one set of node-singular functions is needed. The tetrahedron in (b) has an edge along the feature, so a set of node-singular functions is added for each node on the wedge, along with a set of edge-singular functions for that edge. A tetrahedron on a corner is shown in (c), requiring three sets of node-singular functions and two sets of edge-singular functions. The tetrahedron in (d) requires three sets of node-singular functions and one set of edge-singular functions.}
    \label{fig:sing-tets}
\end{figure}

In an arbitrary geometry, a tetrahedron may have have one or more of its nodes along singular features. At each of these nodes, we introduce ``node-singular" basis functions to represent the corresponding singular behavior. In Fig. \ref{fig:sing-tets}, each of the nodes along a feature is highlighted in red and would include this set of functions. These functions are the tetrahedral extension of those developed for triangles in \cite{2004_Graglia_singular}. This set is ``additive," meaning it is included alongside the standard vector basis functions \cite{2015_Jin_FEM}
\begin{align}
    \mathbf{N}^{ij} = \xi_j \nabla \xi_i - \xi_i \nabla \xi_j,
    \label{eq:std-basis}
\end{align}
where $i,j$ are the indices of the nodes forming the edge and $\xi_i$ is the barycentric coordinate for node $i$. We refer to this function as being ``edge-associated" because its support is the set of tetrahedra sharing edge $ij$. We use the convention of \cite{2015_Jin_FEM} and refer to (\ref{eq:std-basis}) as being a first-order basis function. 

\begin{figure}[t!]
    \centering
    \includegraphics[width=0.95\linewidth]{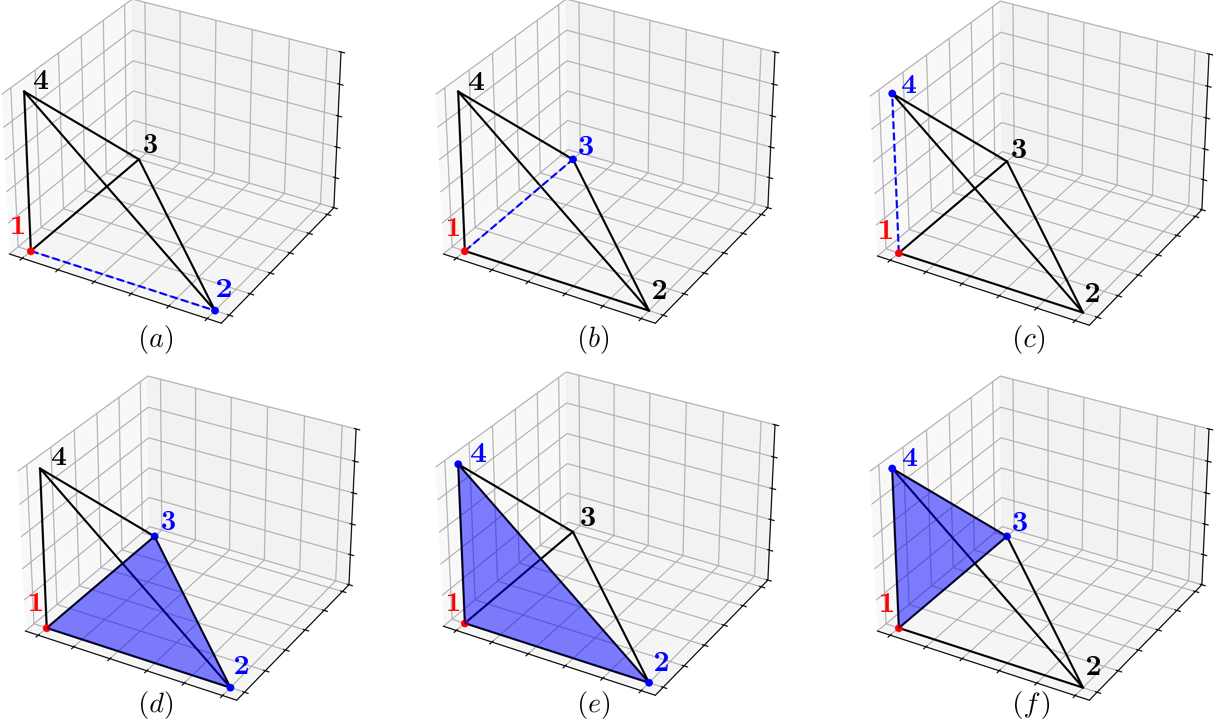}
    \caption{Node-singular basis functions and their geometric associations for a tetrahedron with node $1$ touching a singular feature, as denoted by the red marker. Figs. (a)-(c) show $\mathbf{N}^{ij,g_1}$, where $i = 1$ and $j = 2,3,4$, with their corresponding edges dashed and highlighted in blue. Figs. (d)-(f) show $\mathbf{N}^{ijk,r_1}$, where $i = 1$ and $(j,k) = (2,3), (2,4), (3,4)$, with their associated faces shaded blue.}
    \label{fig:node-basis}
\end{figure}

Our singular set includes functions for representing the gradient-singular electric field in (\ref{eq:sing-E-approx}), which are of the form
\begin{align}
    \mathbf{N}^{ij,g_1} = \nabla \bigg[ \xi_j \big( 1 - (1-\xi_i)^{\nu-1}\big) \bigg],
    \label{eq:sing-point-grad}
\end{align}
and are defined along each edge touching singular node $i$, as shown in Figs. \ref{fig:node-basis}(a)-(c). The superscript $g_1$ is included to indicate that these are first-order gradient functions. Because $1-\xi_i$ can act as a proxy for the radial distance from the singular node within the tetrahedron, these functions represent the $\rho^{\nu-1}$ behavior in (\ref{eq:sing-E-approx}). Further, the functions have a nonzero tangential component only along edge $ij$, allowing them to maintain tangential continuity with neighboring tetrahedra.

More insight into the behavior of these functions can be found by analyzing their behavior along the associated edge. Along edge $ij$, $\xi_i = 1-\xi_j$, so the basis function becomes
\begin{multline}
    \mathbf{N}^{ij,g_1} \bigg |_{\mathrm{edge} \, ij} = \big( 1 - (1-\xi_i)^{\nu-1} \big) \nabla \xi_j \\ + (\nu-1) (1-\xi_i)^{\nu-1} \nabla \xi_i.
\end{multline}
Indexing the two nodes not associated with this edge as $k$ and $\ell$, we note that edge $ij$ is the line where $\xi_k = \xi_\ell = 0$. As a result, the unit vector tangential to the edge can be written as
\begin{align}
    \mathbf{\hat{u}}_{ij} = \frac{\nabla \xi_k \times \nabla \xi_l}{|\nabla \xi_k \times \nabla \xi_l|}.
    \label{eq:tan-dir}
\end{align}
Then, denoting the length of edge $ij$ as $l_{ij}$, we have $\mathbf{\hat{u}}_{ij} \cdot \nabla \xi_i = -1/l_{ij}$ and $\mathbf{\hat{u}}_{ij} \cdot \nabla \xi_j = 1/l_{ij}$. Using these definitions, the tangential component of $\mathbf{N}^{ij,g_1}$ along edge $ij$ is expressed as
\begin{align}
   \mathbf{\hat{u}}_{ij} \cdot \mathbf{N}^{ij,g_1} \bigg |_{\mathrm{edge} \, ij} =  \frac{1 - \nu (1-\xi_i)^{\nu-1}}{l_{ij}},
   \label{eq:sing-point-grad-tan}
\end{align}
giving the desired $\rho^{\nu-1}$ dependence along the edge. The constant part of (\ref{eq:sing-point-grad-tan}) that is not present in (\ref{eq:sing-E-approx}) does not cause any issues because this set of functions is additive, allowing this term to be eliminated in combination with the standard first-order functions of (\ref{eq:std-basis}). More specifically, we see that
\begin{align}
    \mathbf{\hat{u}}_{ij} \cdot \big( -\mathbf{N}^{ij} - \mathbf{N}^{ij,g_1} \big) \bigg |_{\mathrm{edge} \, ij} = \frac{\nu(1-\xi_i)^{\nu-1}}{l_{ij}}.
\end{align}
As shown in Figs. \ref{fig:node-basis}(a)-(c), these functions provide this behavior along all three edges touching the singular node, allowing them to approximate the gradient-singular field of (\ref{eq:sing-E-approx}) throughout the element. Further, the functions have no tangential component at interfaces with nonsingular tetrahedra, so continuity of the tangential component is preserved via the standard basis functions of (\ref{eq:std-basis}) defined along those interfaces.

\begin{figure}[t!]
    \centering
    \includegraphics[width=0.9\linewidth]{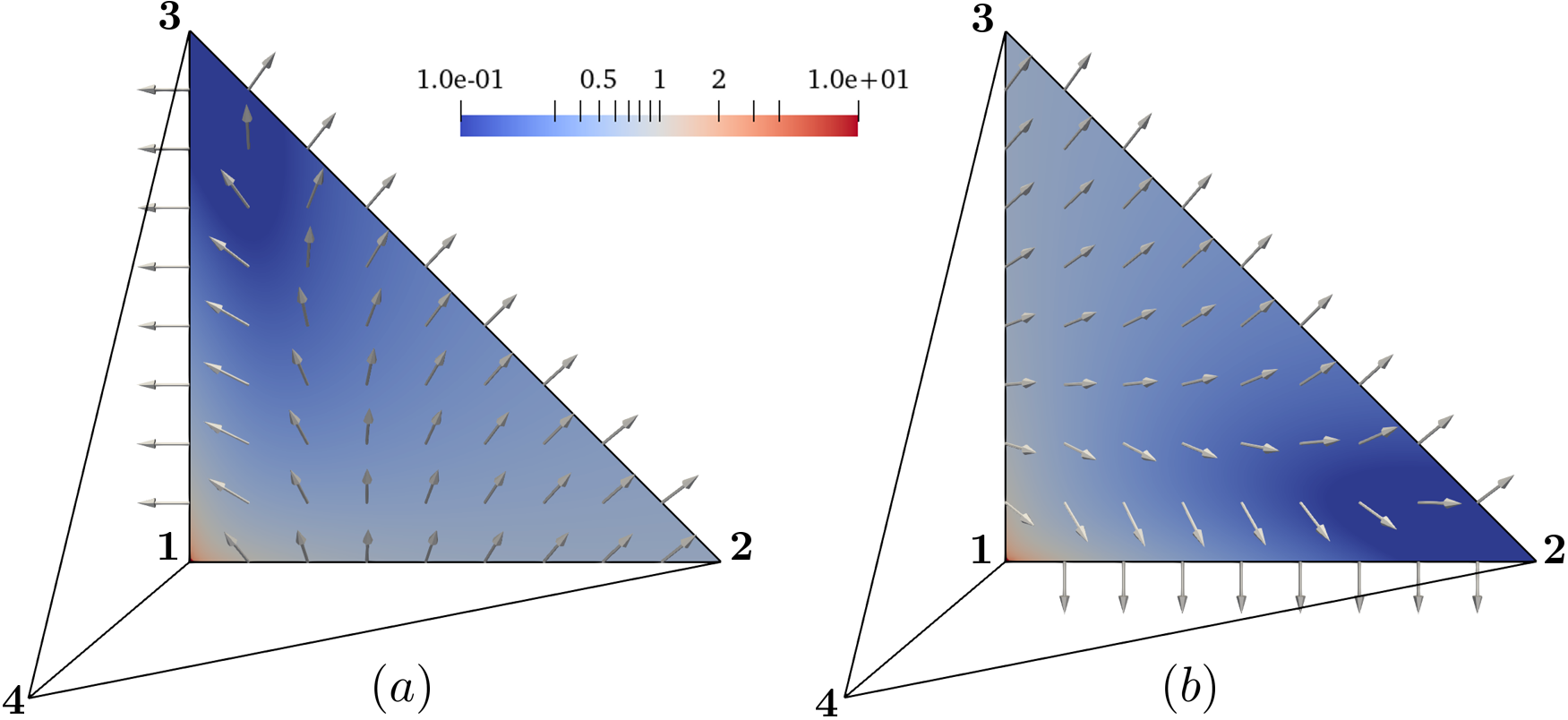}
    \caption{Behavior of (a) $\mathbf{N}^{12,g_1}$ and (b) $\mathbf{N}^{13,g_1}$ with $\nu = 2/3$ along face $123$ of a tetrahedron where node $1$ is touching a singular feature. The color and arrows indicate the magnitude and direction of the function along the face.}
    \label{fig:Nij-g1}
\end{figure}

For further clarity, functions $\mathbf{N}^{12,g_1}$ and $\mathbf{N}^{13,g_1}$ are plotted along face $123$ in Fig. \ref{fig:Nij-g1}. Both functions clearly exhibit singular behavior with respect to node $1$. Additionally, the functions decay to zero as they shift towards the node opposite their associated edges. Each function is normal to the other two edges on the face at points along those edges, which is necessary to maintain tangential continuity. Near their associated edges, they point towards the center of the tetrahedron, allowing the three functions (including $\mathbf{N}^{14,g_1}$) to represent the node-singular behavior throughout the 3D volume of the element.

In \cite{2004_Graglia_singular}, a rotational function is defined in the interior of the triangle to model the magnetic field of (\ref{eq:sing-H-approx}). In a tetrahedron, we add similar functions of the form
\begin{align}
    \mathbf{N}^{ijk,r_1} = \big( (1-\xi_i)^\nu - 1 \big) \mathbf{N}^{jk}
    \label{eq:sing-point-rot}
\end{align}
associated with each of the three faces touching the singular node, as shown in Figs. \ref{fig:node-basis}(d)-(f). In (\ref{eq:sing-point-rot}), $\mathbf{N}^{jk}$ is the standard vector basis function of (\ref{eq:std-basis}). To model the magnetic field properly, the curl of these functions should provide $\rho^\nu$ dependence tangential to the singular feature. In the 2D case in \cite{2004_Graglia_singular}, this can be accomplished using a single function because the tangential direction is assumed to be normal to the 2D geometry being simulated over. In the 3D case considered in this section, the tetrahedron only touches the singular feature at node $i$, so this direction cannot be determined. Instead, we require three functions to allow for the tangential direction to be an arbitrary vector in 3D space.

The utility of (\ref{eq:sing-point-rot}) for providing this behavior can be understood more easily by analyzing its structure normal to the surface the face is associated with. We start by writing the curl of (\ref{eq:sing-point-rot}) as
\begin{multline}
    \nabla \times \mathbf{N}^{ijk,r_1} = -\nu (1 - \xi_i)^{\nu-1} \nabla \xi_i \times \big(\xi_k \nabla \xi_j - \xi_j \nabla \xi_k) \\+ 2 \big( (1 - \xi_i)^\nu - 1 \big) \big( \nabla \xi_k \times \nabla \xi_j \big).
    \label{eq:sing-point-rot-curl}
\end{multline}
The unit vector normal to this surface is $\mathbf{\hat{n}} = \nabla \xi_\ell / | \nabla \xi_\ell |$, so the normal part of the curl includes terms of the form $\nabla \xi_\ell \cdot (\nabla \xi_k \times \nabla \xi_j)$. From \cite[Eq. (45)]{1997_Graglia_interpolatory}, we have
\begin{align}
    \nabla \xi_i \cdot (\nabla \xi_j \times \nabla \xi_k) = \frac{1}{J},
    \label{eq:jacobian}
\end{align}
where $J$ is the Jacobian (which is equal to six times the volume of the tetrahedron). This property holds for cyclic permutations of the indices, and can be extended to include $\nabla \xi_\ell$ using $\nabla \xi_i + \nabla \xi_j + \nabla \xi_k + \nabla \xi_\ell = 0$. More explicitly, we can write
\begin{align}
    \mathbf{\hat{n}} \cdot (\nabla \xi_k \times \nabla \xi_j) = -\frac{\nabla \xi_i + \nabla \xi_j + \nabla \xi_k}{|\nabla \xi_\ell|} \cdot (\nabla \xi_k \times \nabla \xi_j),
\end{align}
where $\nabla \xi_j \cdot (\nabla \xi_k \times \nabla \xi_j) = \nabla \xi_k \cdot (\nabla \xi_k \times \nabla \xi_j) = 0$, so
\begin{align}
    \mathbf{\hat{n}} \cdot (\nabla \xi_k \times \nabla \xi_j) = -\frac{\nabla \xi_i}{|\nabla \xi_\ell|} \cdot (\nabla \xi_k \times \nabla \xi_j) = \frac{1}{J |\nabla \xi_\ell|},
    \label{eq:jaco-1}
\end{align}
which can be repeated for the other terms in (\ref{eq:sing-point-rot-curl}) to obtain
\begin{align}
    \mathbf{\hat{n}} \cdot (\nabla \xi_i \times \nabla \xi_j) = \frac{-1}{J |\nabla \xi_\ell|}, \label{eq:jaco-2} \\
    \mathbf{\hat{n}} \cdot (\nabla \xi_i \times \nabla \xi_k) = \frac{1}{J |\nabla \xi_\ell|}.
    \label{eq:jaco-3}
\end{align}
Then applying (\ref{eq:jaco-1}) through (\ref{eq:jaco-3}) to the normal component of (\ref{eq:sing-point-rot-curl}) along face $ijk$ leads to
\begin{multline}
    \mathbf{\hat{n}} \cdot \nabla \times \mathbf{N}^{ijk,r_1} \bigg|_{\mathrm{face} \, ijk} = \frac{-\nu(1-\xi_i)^{\nu-1}(-\xi_k - \xi_j)}{J |\nabla\xi_\ell|}  \\ + \frac{2 \big( (1-\xi_i)^{\nu} - 1 \big)}{J |\nabla\xi_\ell|}.
\end{multline}
Since $\xi_i + \xi_j + \xi_k = 1$ on face $ijk$, this can be simplified to
\begin{align}
    \mathbf{\hat{n}} \cdot \nabla \times \mathbf{N}^{ijk,r_1} \bigg|_{\mathrm{face} \, ijk} = \frac{-2 + (2+\nu)(1 - \xi_i)^\nu}{J|\nabla\xi_\ell|}.
\end{align}

As in (\ref{eq:sing-point-grad-tan}), there is a constant term which does not pose any issues due to the inclusion of the standard basis functions of (\ref{eq:std-basis}) in this set. In this case, we can write
\begin{align}
    \mathbf{\hat{n}} \cdot \nabla \times \big( \mathbf{N}^{ijk,r_1} + \mathbf{N}^{jk}\big) \bigg|_{\mathrm{face} \, ijk} = \frac{2+\nu}{J |\nabla\xi_\ell|} (1 - \xi_i)^\nu,
\end{align}
indicating that the curl of (\ref{eq:sing-point-rot}) provides the desired $\rho^\nu$ dependence normal to face $ijk$ along that face. Similar results can be shown for the other two functions in Figs. \ref{fig:node-basis}(d)-(f), meaning these basis functions can support this behavior in an arbitrary direction through a linear combination of the three functions.

\subsubsection{Edge-Singular Functions}
\label{subsubsec:edge-funcs}

\begin{figure}
    \centering
    \includegraphics[width=0.75\linewidth]{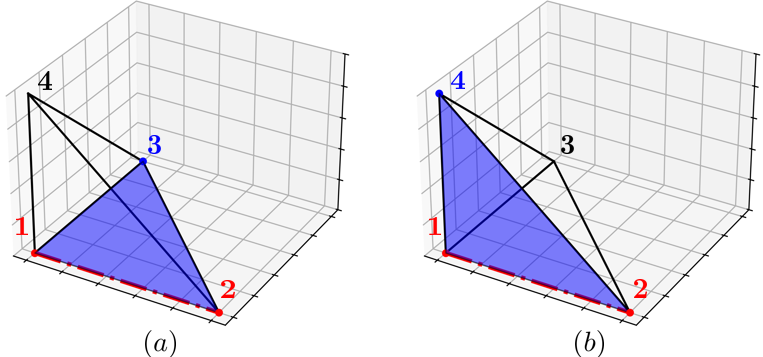}
    \caption{Edge-singular basis functions and their geometric associations for a tetrahedron with edge $12$ along a singular feature, as denoted by the red dash-dotted line. The functions $\mathbf{N}^{123,g_1}$ and $\mathbf{N}^{124,g_1}$ have their associated faces shaded blue in (a) and (b) respectively.}
    \label{fig:edge-basis}
\end{figure}

The functions in Section \ref{subsubsec:node-funcs} exhibit singular behavior with respect to a single node of a tetrahedron. This means that when used in isolation, these functions provide poor accuracy in modeling fields within tetrahedra with an edge directly along a singular feature. To resolve this issue, we supplement them with additional ``edge-singular" functions for modeling this behavior. To ensure that these functions have zero tangential component at interfaces with tetrahedra which do not share the edge, we associate them with the two faces touching singular edge $ij$, as shown in Fig. \ref{fig:edge-basis}. These functions are expressed as
\begin{align}
    \mathbf{N}^{ijk,g_1} = \nabla \bigg[ \xi_i \xi_j \xi_k \big( 1 - (1-\xi_i-\xi_j)^{\nu-1}\big) \bigg].
    \label{eq:sing-edge-grad}
\end{align}
These functions are structured to provide the same general properties as (\ref{eq:sing-point-grad}), but using $1 - \xi_i - \xi_j$ as the proxy for $\rho$ so they are singular with respect to the entire edge. However, making this change to (\ref{eq:sing-point-grad}) in isolation would violate tangential continuity with adjacent elements. To correct this, an additional $\xi_i \xi_j$ factor is included in (\ref{eq:sing-edge-grad}) to ensure the functions have zero tangential component along all edges and faces besides face $ijk$. Therefore, the functions can model edge-singular behavior while maintaining continuity with neighboring elements.

To see how these functions provide the needed behavior, it is helpful to examine their properties along the face they are associated with. For instance, along face $ijk$ the component of (\ref{eq:sing-edge-grad}) tangential to the face can be approximated by
\begin{multline}
    \mathbf{\hat{n}} \times \mathbf{\hat{n}} \times \mathbf{N}^{ijk,g_1} \bigg|_{\mathrm{face} \, ijk} \approx \big( 1-\nu (1 - \xi_i - \xi_j)^{\nu-1} \big) \\ \times \xi_i \xi_j \big( \mathbf{\hat{n}} \times \mathbf{\hat{n}} \times \nabla \xi_k),
    \label{eq:sing-edge-grad-tan}
\end{multline}
where the steps to derive (\ref{eq:sing-edge-grad-tan}) from (\ref{eq:sing-edge-grad}) are outlined in Appendix \ref{app:edge-sing}. Here, we see there is a constant term in the first set of parentheses on the right-hand side that is then multiplied by a $\xi_i \xi_j$ factor, which is needed to maintain tangential continuity. In this case, this non-ideal term cannot be canceled out using the standard first-order basis functions of (\ref{eq:std-basis}). However, this term can still be eliminated if standard higher-order basis functions are used, allowing (\ref{eq:sing-edge-grad-tan}) to exhibit the desired $\rho^{\nu-1}$ behavior.

In Section \ref{subsubsec:node-funcs}, three gradient functions were needed to represent the gradient-singular field throughout the element. While this is necessary when only a single node is touching the singular feature, when an edge is along the feature we can take advantage of (\ref{eq:sing-E-approx}) having no singular component tangential to the edge. Within the tetrahedron, $\nabla \xi_k$ and $\nabla \xi_\ell$ form a basis for the set of directions orthogonal to the direction of the edge, as described by (\ref{eq:tan-dir}). Then we observe from (\ref{eq:sing-edge-grad-tan}) that the two functions shown in Fig. \ref{fig:edge-basis} approximate the desired $\rho^{\nu-1}$ behavior in each of these directions, allowing them to represent the gradient-singular field of (\ref{eq:sing-E-approx}).

\begin{figure}[t!]
    \centering
    \includegraphics[width=0.9\linewidth]{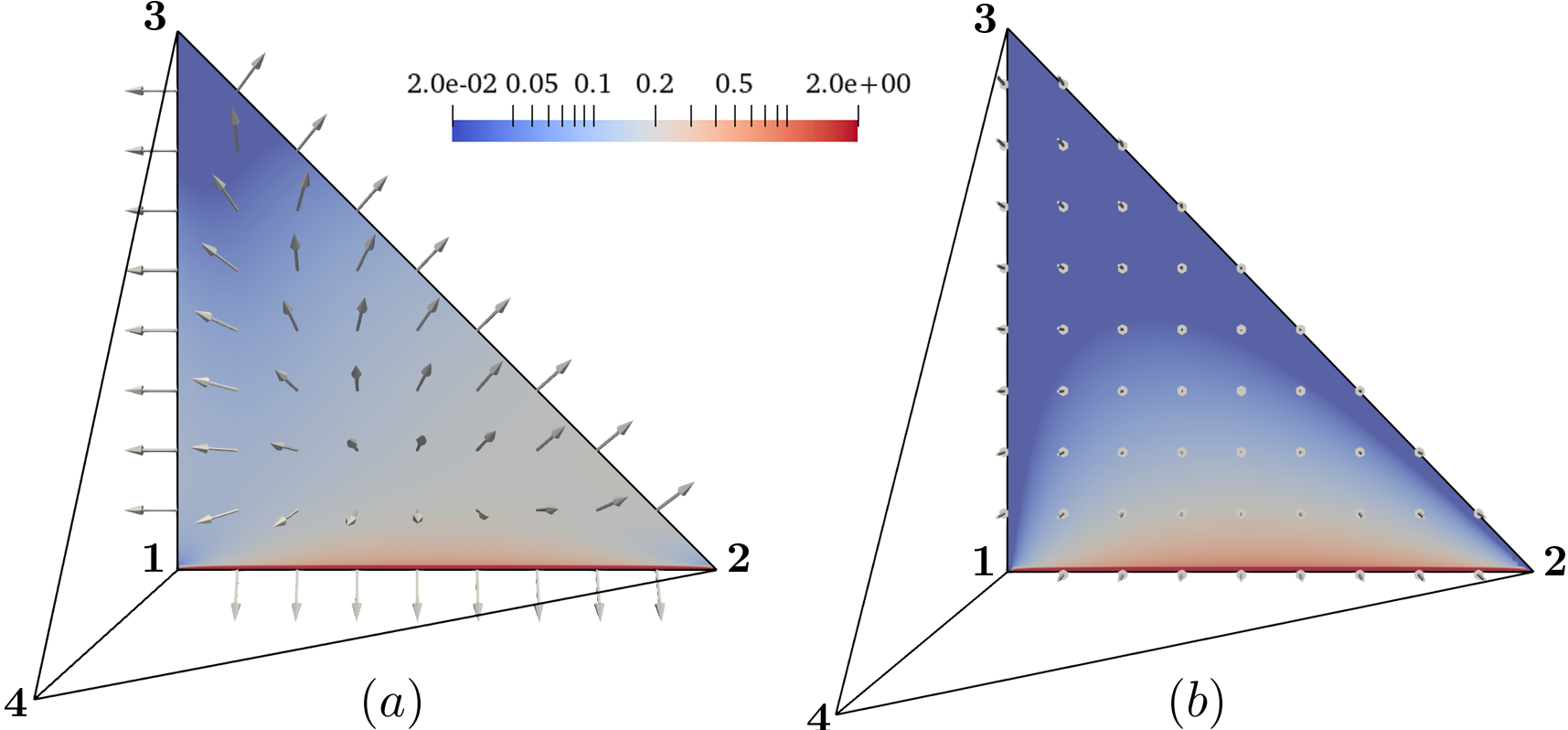}
    \caption{Behavior of (a) $\mathbf{N}^{123,g_1}$ and (b) $\mathbf{N}^{124,g_1}$ with $\nu = 2/3$ along face $123$ of a tetrahedron with edge $12$ along a singular feature. The color and arrows indicate the magnitude and direction of the function along the face.}
    \label{fig:Nijk-g1}
\end{figure}

The key features of these functions are visualized in Fig. \ref{fig:Nijk-g1}, where $\mathbf{N}^{123,g_1}$ and $\mathbf{N}^{124,g_1}$ are plotted along face $123$. Both functions are clearly singular with respect to edge $12$, and their magnitude dissipates very close to nodes $1$ and $2$ due to the $\xi_i \xi_j$ factor. This behavior is desirable because these functions are included alongside the node-singular functions of (\ref{eq:sing-point-grad}), meaning the singular behavior with respect to those nodes is already accounted for by those functions. In Fig. \ref{fig:Nijk-g1}(a), we see that $\mathbf{N}^{123,g_1}$ has a nonzero component tangential to the face, but is normal to all three edges bounding the face at points on the edges to satisfy tangential continuity. Further, because $\mathbf{N}^{124,g_1}$ can only have a nonzero tangential component along face $124$, it is fully normal to face $123$ in Fig. \ref{fig:Nijk-g1}(b).

Similarly to the set of node-singular functions, we can define rotational functions to model the magnetic field of (\ref{eq:sing-H-approx}) in the vicinity of the edge. In this case, we introduce a single function associated with the volume and of the form
\begin{align}
    \mathbf{N}^{ijk\ell,r_1} = \big( (1-\xi_i - \xi_j)^\nu - 1 \big) \xi_i \xi_j \mathbf{N}^{k\ell}.
    \label{eq:sing-edge-rot}
\end{align}
Once again, we include a $\xi_i \xi_j$ factor to force the tangential component of the function to zero along all edges and faces of the tetrahedron, and ensure continuity with adjacent elements.

As described in Section \ref{subsubsec:node-funcs}, three rotational functions are needed to represent the magnetic field when the direction tangential to the singular feature is unknown. But since this direction is known when the tetrahedron has an edge along the feature, only a single function with curl parallel to the edge is needed. In Appendix \ref{app:edge-sing-rot}, we demonstrate that the portion of the curl of (\ref{eq:sing-edge-rot}) tangential to edge $ij$ can be approximated as
\begin{multline}
    \mathbf{\hat{u}}_{ij} \cdot \nabla \times \mathbf{N}^{ijk\ell,r_1} \approx \big( 2 - (2 + \nu) (1-\xi_i - \xi_j)^\nu \big) \\
    \times \xi_i \xi_j |\nabla \xi_k \times \nabla \xi_\ell|.
    \label{eq:sing-edge-rot-curl-approx}
\end{multline}
So, like (\ref{eq:sing-point-rot}), the curl of (\ref{eq:sing-edge-rot}) provides the desired $\rho^{\nu}$ behavior tangential to the singular feature, with the weighting factor $\xi_i \xi_j$ included in this case to enforce the tangential continuity. Similarly as for (\ref{eq:sing-edge-grad-tan}), (\ref{eq:sing-edge-rot-curl-approx}) also contains a non-ideal constant term in the first set of parentheses on the right-hand side that can only be eliminated in combination with standard higher-order basis functions due to the $\xi_i \xi_j$ weighting factor.

\subsection{Higher-Order Basis Functions}
\label{subsec:higher-order}

\subsubsection{Node-Singular Functions}
\label{subsubsec:node-funcs-high}

As in \cite{2004_Graglia_singular}, we derive higher-order interpolatory versions of (\ref{eq:sing-point-grad}) and (\ref{eq:sing-point-rot}) using Silvester-Lagrange interpolating polynomials, which are described in detail in \cite{1997_Graglia_interpolatory}. This includes the interpolating polynomials
\begin{align}
    R_a^{s+1}(\xi_i) = \frac{1}{a!} \prod_{m=0}^{a-1} \big[ (s+1) \xi_i - m\big],
    \label{eq:lagrange}
\end{align}
which are structured such that for integer $0 \leq n \leq a \leq s-1$, 
\begin{align}
    R_a^{s+1}(\tfrac{n}{s+1}) = 
    \begin{cases}
        0, & n < a \\
        1, & n = a
    \end{cases} \,.
    \label{eq:lagrange-cases}
\end{align}
Additionally, we have shifted interpolating polynomials
\begin{align}
    \hat{R}_a^{s+1}(\xi_i) = \frac{1}{(a-1)!} \prod_{m=1}^{a-1} \big[ (s+1) \xi_i - m\big],
    \label{eq:lagrange-shifted}
\end{align}
which also provide the same behavior as (\ref{eq:lagrange-cases}), but for $1 \leq n \leq a \leq s$. Here, $s$ denotes the order of the functions.

\begin{figure}[t]
    \centering
    \includegraphics[width=0.9\linewidth]{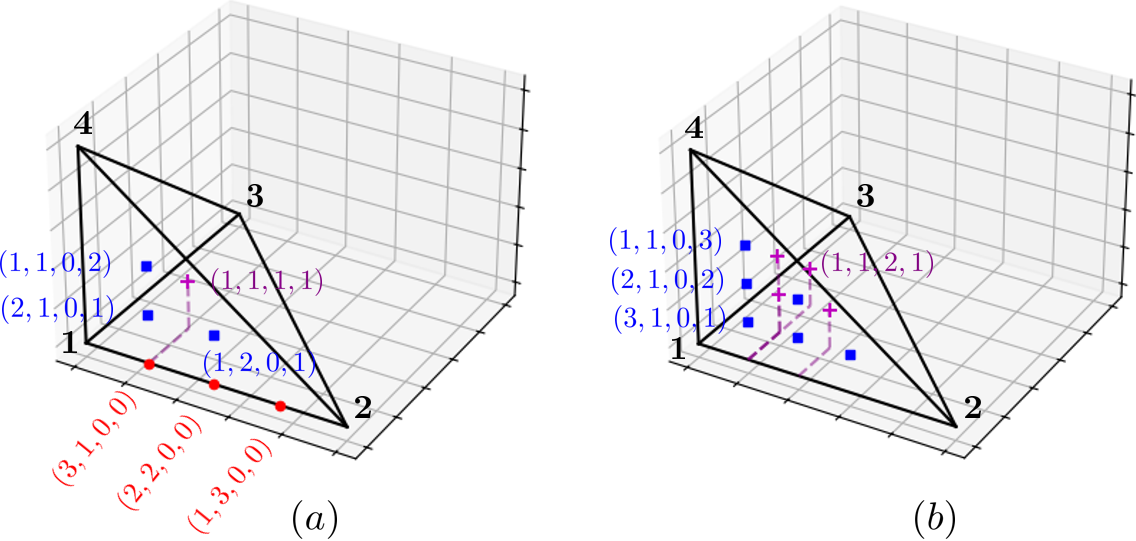}
    \caption{Interpolation points for higher order basis functions. In (a), points for a third-order edge-associated function on edge $12$ are shown, where all points satisfy $a+b+c+d = s+1$. Points along the edge with $a,b \neq 0$ are shown with red dots, points on face $ijk$ with $a,b,c \neq 0$ are indicated by blue squares, and a purple `+' is used for the point in the volume. Points on face $ijk$ are omitted for visual clarity. Points for a third-order face-associated function on face $ij\ell$ are shown in (b), where $a+b+c+d=s+2$.}
    \label{fig:high-order}
\end{figure}

These polynomials are used to define a grid of ``interpolation points," as pictured in Fig. \ref{fig:high-order}. Each point is defined by indices $(a,b,c,d)$, where the interpolating function for that point is of the general form $R_a^{s+1}(\xi_i) R_b^{s+1}(\xi_j) R_c^{s+1}(\xi_k) R_d^{s+1}(\xi_\ell)$. Due to the structure in (\ref{eq:lagrange-cases}), each interpolating function will be nonzero at its associated grid point, and zero at all other grid points. If only the non-shifted polynomials of (\ref{eq:lagrange}) are used, this grid would include points on the nodes of the tetrahedron. However, the basis functions used here are designed to be tangentially continuous along specified interfaces, such as edges for the functions of (\ref{eq:sing-point-grad}). In these cases, we wish to shift the grid away from the other interfaces in the tetrahedron to avoid violating continuity there. This is accomplished using the shifted functions of (\ref{eq:lagrange-shifted}), which move the grid away from the node associated with the function. This is demonstrated in Fig. \ref{fig:high-order}(a) for a function associated with edge $12$. The grid is shifted away from those nodes to avoid having interpolation points on other edges or faces not touching edge $12$. 


The interpolating polynomials provide a means of generating basis functions spanning higher-order polynomial spaces. For the singular basis functions, this means the higher-order functions can better represent the nonsingular terms in the Meixner expansion of (\ref{eq:sing-E}) and (\ref{eq:sing-H}) while retaining the underlying singular behavior and tangential continuity of the first-order functions. The order of the singular functions can be selected independently from the standard basis functions, so a pair of indices $(p, s)$ are used to enumerate the orders of standard and singular basis functions, respectively. 

For the edge-based functions of (\ref{eq:sing-point-grad}), we use the shifted polynomials of (\ref{eq:lagrange-shifted}) for the nodes forming the edge to avoid interpolating where the function has no tangential component. Indexing the nodes opposite the edge as $k,\ell$ (see nodes $3,4$ in Fig. \ref{fig:node-basis}(a)), the order $s$  gradient-singular functions are
\begin{multline}
    \mathbf{N}^{ij,g_s}_{abcd} = \nabla \bigg[\hat{R}_a^{s+1}(\xi_i) \hat{R}_b^{s+1} (\xi_j) R_c^{s+1}(\xi_k) R_d^{s+1}(\xi_\ell) \\ \times \xi_j \big( 1 - (1-\xi_i)^{\nu-1}\big) \bigg],
    \label{eq:sing-point-grad-s}
\end{multline}
with shifted interpolation point indices $a, b \in [1,s]$, the remaining point indices $c, d \in [0,s-1]$, and $a+b+c+d = s+1$. The interpolation polynomials are included inside the gradient to make the higher-order functions irrotational. As indicated by Fig. \ref{fig:high-order}(a), the interpolation points for (\ref{eq:sing-point-grad-s}) may lie on edge $ij$, on a face touching the edge, or within the tetrahedron volume.

As for the standard higher-order basis functions \cite{1997_Graglia_interpolatory}, there are dependency relations in the functions of (\ref{eq:sing-point-grad-s}) which require some functions to be eliminated to produce a full-rank system. Only the functions defined on an edge have interpolation points along it, so none of the $s$ points there need to be eliminated. For edge $ij$, this is the set of points $(a,b,0,0)$. 

The interpolation points on each face include functions from each edge touching a singular node. If node $i$ is singular, we have functions $\mathbf{N}_{abc0}^{ij,g_s}$ and $\mathbf{N}_{abc0}^{ik,g_s}$ at each point $(a,b,c,0)$ with $a,b,c \neq 0$ on face $ijk$. However, these functions are linearly dependent, meaning one set must be removed. This is straightforward to show for the second-order case, where
\begin{align}
    \mathbf{N}^{ij,g_2}_{1110} = \mathbf{N}^{ik,g_2}_{1110} = 3 \nabla \bigg[ \xi_j \xi_k \big( 1 - (1-\xi_i)^{\nu-1}\big)\bigg],
    \label{eq:sing-point-grad-dependence}
\end{align}
so one function needs to be eliminated. In general, there is only one independent function in the form of (\ref{eq:sing-point-grad-s}) for each of the $s(s-1)/2$ grid points on the face. This logic can be repeated for points on faces $ij\ell$ and $ik\ell$ touching the singular nodes, giving a total of $3s(s-1)/2$ functions for interpolation points on faces. 

The functions at points inside the volume with $a,b,c,d \neq 0$ have the same dependency relation as for the faces. For example, in the third-order case we have $\mathbf{N}^{ij,g_3}_{1111} = \mathbf{N}^{ik,g_3}_{1111} = \mathbf{N}^{i\ell,g_3}_{1111}$. As a result, we eliminate all but one function for each of the $s(s-1)(s-2)/6$ internal points.

The procedure for constructing higher-order rotational functions from (\ref{eq:sing-point-rot}) is similar to the prior functions, but shifted interpolating polynomials are needed for all nodes on the associated face. Otherwise, we would construct higher-order functions defined on one face with interpolation points along the other faces, where the first-order function has zero tangential component. For example, the function $N^{ijk,g_2}_{0111}$ would be defined along face $ijk$, but the interpolation point is on face $jk\ell$, leading to continuity violations. To correct this, we use
\begin{align}
    \mathbf{N}_{abcd}^{ijk,r_s} = \hat{R}_a^{s+2}(\xi_i) \hat{R}_b^{s+2} (\xi_j) \hat{R}_c^{s+2}(\xi_k) R_d^{s+2}(\xi_\ell) \mathbf{N}^{ijk,r_1},
    \label{eq:sing-point-rot-s}
\end{align}
where $a,b,c \in [1,s]$ and $d \in [0,s-1]$, with $a+b+c+d = s+2$. Because three polynomials are shifted instead of two in (\ref{eq:sing-point-grad-s}), we use interpolatory polynomials $\hat{R}_a^{s+2}$ in (\ref{eq:sing-point-rot-s}) rather than $\hat{R}_a^{s+1}$ to ensure the functions interpolate to evenly spaced grid points, as shown in Fig. \ref{fig:high-order}(b). 

Using three shifted polynomials also means there are no interpolation points along edges due to the requirement $a,b,c > 0$. Additionally, the interpolation points on face $ijk$ only include functions $\mathbf{N}^{ijk,r_1}_{abc0}$, meaning there is one function for each of the $(s+1)s/2$ points on the face, and none are eliminated. Within the volume, we can show that there are dependencies between the functions defined on different faces, as we have for the standard basis functions
\begin{align}
    \xi_\ell\mathbf{N}^{jk} - \xi_k\mathbf{N}^{j\ell} + \xi_j \mathbf{N}^{k\ell} = 0.
\end{align}
To eliminate this dependency we remove one set of functions in the volume, leaving two functions for each of the $(s+1)s(s-1)/6$ internal points. The number of basis functions associated with each piece of geometry are given in Table \ref{table:basis-count}.

\begin{table}[t]
\caption{Number of order $s$ singular basis functions associated with each piece of tetrahedron geometry} 
\renewcommand{\arraystretch}{1.5}
\centering
\newcolumntype{D}{>{\centering\arraybackslash} m{0.09\textwidth}}
\newcolumntype{M}{>{\centering\arraybackslash} m{0.11\textwidth}}
\footnotesize
\label{table:basis-count}
\begin{tabular}{D|D|D|M}
 \hline
 \hline
 \textbf{Basis Function} & \textbf{Edge Functions} & \textbf{Face Functions} & \textbf{Volume Functions} \\
 \hline
 \hline
 (\ref{eq:sing-point-grad-s}) & $3 \,\mathrm{edges} \times s$ & $3 \, \mathrm{faces} \times s(s-1)/2$ & $s(s-1)(s-2)/6$\\
 \hline
 (\ref{eq:sing-point-rot-s}) & $-$ & $3 \, \mathrm{faces} \times (s+1)s/2$ & $(s+1)s(s-1)/3$\\
 \hline
 (\ref{eq:sing-edge-grad-s}) & $-$ & $2 \, \mathrm{faces} \times (s+1)s/2$ & $(s+1)s(s-1)/6$\\
 \hline
 (\ref{eq:sing-edge-rot-s}) & $-$ & $-$ & $(s+2)(s+1)s/6$ \\
 \hline
 \hline
\end{tabular}
\end{table}

\subsubsection{Edge-Singular Functions}
\label{subsubsec:edge-funcs-high}

Higher-order versions of (\ref{eq:sing-edge-grad}) and (\ref{eq:sing-edge-rot}) can be derived via the same process as in Section \ref{subsubsec:node-funcs-high}. For (\ref{eq:sing-edge-grad}), shifted interpolation functions are used for all three nodes along face $ijk$, so we obtain
\begin{multline}
    \mathbf{N}^{ijk,g_s}_{abcd} = \nabla \bigg[\hat{R}_a^{s+2}(\xi_i) \hat{R}_b^{s+2} (\xi_j) \hat{R}_c^{s+2}(\xi_k) R_d^{s+2}(\xi_\ell) \\ \times \xi_i \xi_j \xi_k \big( 1 - (1-\xi_i-\xi_j)^{\nu-1}\big) \bigg],
    \label{eq:sing-edge-grad-s}
\end{multline}
where like (\ref{eq:sing-point-rot-s}),  $a,b,c \in [1,s]$, $d \in [0,s-1]$, and $a+b+c+d = s+2$. There are once again no interpolation points on edges, and there is one function for each of the $(s+1)s/2$ points on the faces touching the singular edge. The functions in the volume can be shown to be dependent by the same process as (\ref{eq:sing-point-grad-dependence}), meaning only one of the two functions is included for each of the $(s+1)s(s-1)/6$ points in the volume.

Because (\ref{eq:sing-edge-rot}) is defined within the volume, we use shifted interpolating polynomials for all four nodes, leading to
\begin{align}
    \mathbf{N}_{abcd}^{ijk\ell,r_s} = \hat{R}_a^{s+3}(\xi_i) \hat{R}_b^{s+3} (\xi_j) \hat{R}_c^{s+3}(\xi_k) \hat{R}_d^{s+3}(\xi_\ell) \mathbf{N}^{ijk\ell,r_1},
    \label{eq:sing-edge-rot-s}
\end{align}
with $a,b,c,d \in [1,s]$ and $a+b+c+d = s+3$. In this case, all $(s+2)(s+1)s/6$ points are within the volume, and no functions are eliminated. The number of functions associated with each piece of geometry are given in Table \ref{table:basis-count}.

\subsection{Implementation Details}
\label{subsec:implementation}

In this section, we comment on some of the practical aspects of implementing these functions in an efficient manner. As shown in Fig. \ref{fig:sing-tets}, tetrahedra in realistic 3D geometries may contact singular features along several edges and nodes. As discussed previously, this is accounted for by repeating the sets of node- and edge-singular functions for each singular feature touched by the element. This approach ensures tangential continuity is maintained between tetrahedra in these regions without requiring excessively complex basis functions. 

A drawback of this approach is that the number of functions can become quite large as the order is increased, which among other factors can increase the matrix assembly time. However, this can be largely mitigated by pre-calculating finite element matrices for a unit tetrahedron, and simply rescaling them for each tetrahedron in the mesh \cite{1999_Webb_Hierarchical}. Since our basis functions are designed to model the electric field, we solve the vector Helmholtz equation for $\mathbf{E}$, which in discretized form is \cite{2015_Jin_FEM}
\begin{align}
    [S] \{ e \} - \omega^2[T] \{ e \} = 0,
    \label{eq:wave-discretized}
\end{align}
where $\mathbf{E} = \sum_n \mathbf{N}_n \{e\}_n$, $\mathbf{N}_n$ refers to an arbitrary basis function with index $n$, and $\{e\}_n$ is the associated expansion coefficient. In (\ref{eq:wave-discretized}), we assume a source-free and lossless simulation domain for simplicity, but these terms can easily be included. The $(m, n)$ entry of the mass matrix $[T]$ is
\begin{align}
    [T]_{mn} = \int_\Omega \epsilon \mathbf{N}_m \cdot \mathbf{N}_n \, \mathrm{d}V = \sum_t \int_{\Omega_t} \epsilon \mathbf{N}_m \cdot \mathbf{N}_n \, \mathrm{d}V,
    \label{eq:mass-matrix}
\end{align}
where $\Omega$ is the entire simulation domain, and $\Omega_t$ is the spatial support of tetrahedron $t$. Similarly, for the stiffness matrix $[S]$ the $(m,n)$ entry is given by
\begin{align}
    [S]_{mn} = \sum_t \int_{\Omega_t} \mu^{-1} (\nabla \times \mathbf{N}_m) \cdot (\nabla \times \mathbf{N}_n) \, \mathrm{d}V.
    \label{eq:stiff-matrix}
\end{align}

Rather than evaluating (\ref{eq:mass-matrix}) and (\ref{eq:stiff-matrix}) directly in each tetrahedron, we can instead move the portions of the basis functions dependent on the geometry of the tetrahedron outside the integral, allowing it to be precalculated and saved. For (\ref{eq:mass-matrix}), this can be done by rewriting each basis function in the form
\begin{align}
    \mathbf{N}_m = \sum_{u \in i,j,k} N_{m,u} \nabla \xi_u,
    \label{eq:basis-sub}
\end{align}
where any basis function can be expressed in this form using $\xi_i + \xi_j + \xi_k + \xi_l = 1$. Then (\ref{eq:mass-matrix}) can be expressed as
\begin{align}
    [T]_{mn} = \sum_t \epsilon \sum_{u \in i,j,k} \sum_{v \in i,j,k} \nabla \xi_{u,t} \cdot \nabla \xi_{v,t} \int_{\Omega_t} N_{m,u} N_{n,v} \, \mathrm{d}V,
\end{align}
where $\nabla \xi_{u,t}$ is dependent on the tetrahedron geometry. The integral can be written in barycentric coordinates so that
\begin{multline}
    [T]_{mn} = \sum_t 6 V_t \epsilon \sum_{u \in i,j,k} \sum_{v \in i,j,k} \nabla \xi_{u,t} \cdot \nabla \xi_{v,t} \\ \times \int_0^1 \int_0^{1-\xi_i} \int_0^{1-\xi_i-\xi_j} N_{m,u} N_{n,v} \, \mathrm{d}\xi_k \mathrm{d}\xi_j \mathrm{d}\xi_i,
    \label{eq:mass-bary}
\end{multline}
where $V_t$ is the volume of tetrahedron $t$. The integral in (\ref{eq:mass-bary}) is fully independent of the geometry of a tetrahedron since $N_{m,u}$ and $N_{n,v}$ only depend on the barycentric coordinates themselves. As a result, these integrals can be pre-calculated for every combination of basis functions and saved to a file. The contributions of each tetrahedron to $[T]$ can then be calculated inexpensively through weighted sums of these integrals scaled by $\nabla \xi_{u,t}$ and $V_t$. 

\begin{figure}[t!]
    \centering
    \includegraphics[width=0.8\linewidth]{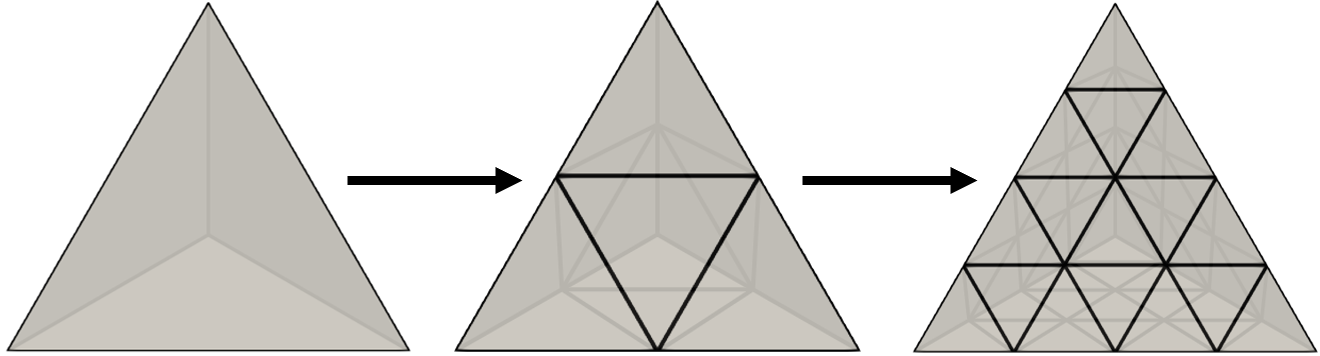}
    \caption{Recursive splitting process used to improve numerical integration accuracy. The unit tetrahedron is split into $8$ smaller tetrahedra on the first pass, and each of these is split again to get $64$ tetrahedra on the second pass. Each successive pass increases the number of tetrahedra by a factor of $8$.}
    \label{fig:tet-splits}
\end{figure}

A similar process can be performed to assemble the stiffness matrix in (\ref{eq:stiff-matrix}). In this case, we can write the curl in the form
\begin{align}
    \nabla \times \mathbf{N}_m = \sum_{u \in i,j,k} C_{m,u} \mathbf{e}_u,
\end{align}
where $\mathbf{e}_i = \nabla \xi_j \times \nabla \xi_k$, $\mathbf{e}_j = \nabla \xi_k \times \nabla \xi_i$, and $\mathbf{e}_k = \nabla \xi_i \times \nabla \xi_j$. Following the same approach as for $[T]$ leads to the expression
\begin{multline}
    [S]_{mn} = \sum_t 6 V_t \mu^{-1} \sum_{u \in i,j,k} \sum_{v \in i,j,k} \mathbf{e}_{u,t} \cdot \mathbf{e}_{v,t} \\ \times \int_0^1 \int_0^{1-\xi_i} \int_0^{1-\xi_i-\xi_j} C_{m,u} C_{n,v} \, \mathrm{d}\xi_k \mathrm{d}\xi_j \mathrm{d}\xi_i,
    \label{eq:stiff-bary}
\end{multline}
where the integrals can again be pre-calculated to accelerate matrix assembly.

In our implementation, the basis functions are calculated numerically on a unit tetrahedron where $\nabla \xi_i = \mathbf{\hat{x}}$, $\nabla \xi_j = \mathbf{\hat{y}}$, and $\nabla \xi_k = \mathbf{\hat{z}}$, so the functions are naturally in the form of (\ref{eq:basis-sub}). This process is performed for all singular functions from all nodes or edges, allowing it to be applied to any tetrahedron in the mesh. We perform the integrals numerically using an $8$th order Gaussian quadrature rule for tetrahedra \cite{1986_Keast_Quadrature}. To further improve accuracy, we refine the tetrahedron by recursively splitting it about the midpoints of each of its edges into $8$ smaller tetrahedra, then sum the integrals over each sub-element. The recursive splitting procedure is captured visually in Fig. \ref{fig:tet-splits}. We repeated this splitting process six times, which we found to produce accurate simulation results throughout the tests discussed in Section \ref{sec:results}.

\section{Results}
\label{sec:results}
In this section, we validate the singular basis functions and demonstrate their computational benefits. First, we compare to analytical solutions on a grounded coplanar waveguide geometry in Section \ref{subsec:val-ana}. Subsequently, in Section \ref{subsec:val-num}, we compare the field behavior produced by our functions near singular features to that of standard basis functions using an extremely fine mesh. These examples are used to confirm that these singular basis functions are representing the correct underlying field behavior. Finally, we simulate a superconducting qubit geometry in Section \ref{subsec:part-ratios} and compare the resulting participation ratios to those extracted from Ansys HFSS to demonstrate the efficiency improvements gained from using our introduced singular basis functions.

\subsection{Analytical Validation}
\label{subsec:val-ana}

\begin{figure}[t!]
    \centering
    \includegraphics[width=0.9\linewidth]{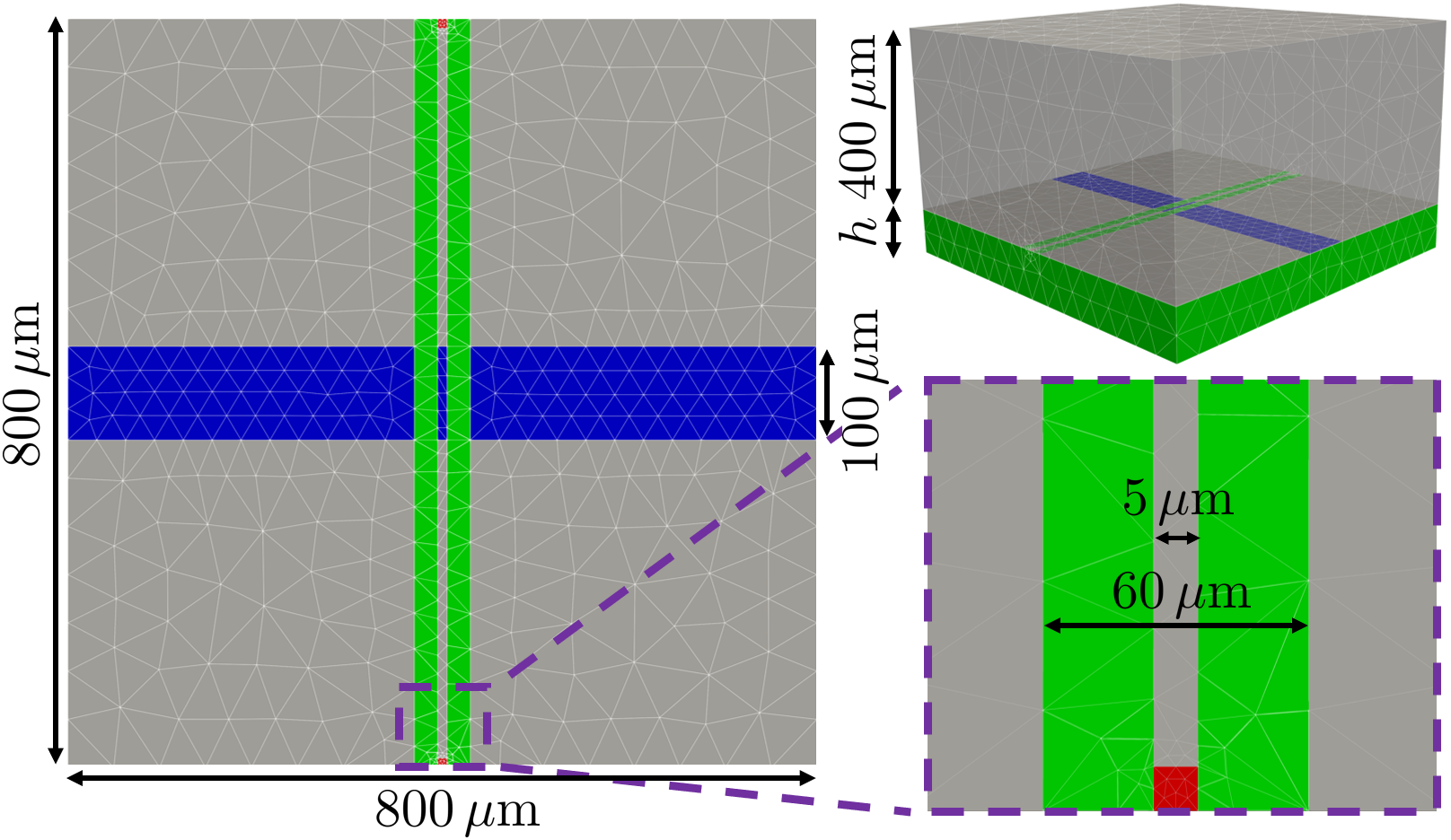}
    \caption{The grounded coplanar waveguide geometry. The gray regions are the conductors, which all have zero thickness. The exposed silicon substrate ($\epsilon_r = 11.9$) is highlighted green, while the ports for driving and terminating the waveguide are in red. The metal-substrate participation ratio is calculated only over the central $100 \, \mu \mathrm{m}$ region, which is colored blue. The edges of the planar geometry are flush with a perfectly conducting bounding box which extends $400 \, \mu \mathrm{m}$ above the pictured plane. The substrate extends a height $h$ below this plane, and its bottom and sides are along the boundary of the box.}
    \label{fig:gcpw}
\end{figure}

We begin by extracting the per-unit-length (PUL) capacitance of the grounded coplanar waveguide (GCPW) geometry considered in \cite{2026_Wang_SIE} and shown in Fig. \ref{fig:gcpw}. To calculate the PUL capacitance, we drive the device with a $1 \, \mathrm{A}$ current at the input port with both ports open-circuited and measure the voltage at the input to calculate the input impedance $Z_\mathrm{in}$. This simulation is ran at $1 \, \mathrm{GHz}$ to ensure that the geometry is dominantly capacitive, allowing the PUL capacitance to be approximated by $C \approx 1/(j \omega Z_\mathrm{in} \ell)$, where $\ell$ is the length of the waveguide.

\begin{figure}[t!]
    \centering
    \includegraphics[width=0.8\linewidth]{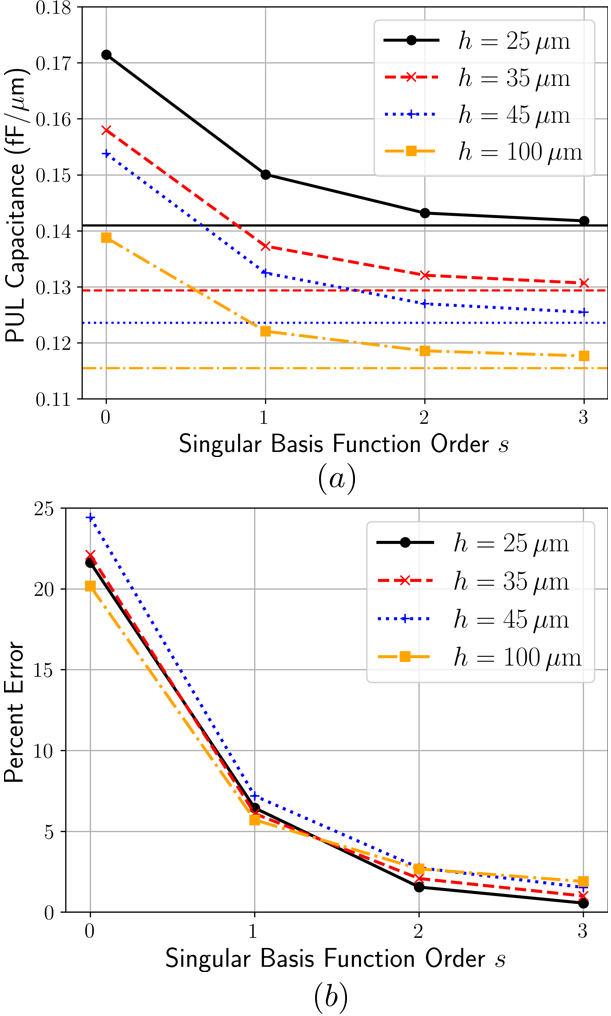}
    \caption{Capacitance of the GCPW with varying substrate height $h$. The order of the standard basis functions is fixed at $p = 2$ while the order of the singular functions is swept. The PUL capacitance is plotted in (a) with the analytical solutions shown by the horizontal lines. The percent error between the numerical and analytical results is shown in (b).}
    \label{fig:GCPW-cap}
\end{figure}

In Fig. \ref{fig:GCPW-cap}(a), we plot the PUL capacitance of the GCPW with varying substrate height $h$. The $x$-axis gives the order of the singular basis functions, where $s=0$ indicates that no singular functions are used. For all of these results, the standard basis functions are order $p=2$. Since for these simulations the GCPW conductors have zero thickness, when the singular basis functions are included we treat all conductive edges in the interior of the geometry as singular with $\nu = 0.5$ to match the expected behavior near these features \cite{1972_Meixner_Singular}. This causes tetrahedra with faces touching the signal trace to touch multiple singular features, but this is accommodated for by using multiple sets of functions as described in Section \ref{subsec:implementation}.

The horizontal lines in Fig. \ref{fig:GCPW-cap}(a) show the analytical solutions from \cite{1987_Ghione_GCPW, 2026_Wang_SIE} at each height, with the percent error between the solutions plotted in Fig. \ref{fig:GCPW-cap}(b). In general, we see very high error with no singular functions ($s=0$), and a very strong improvement in accuracy using even only the first-order singular functions. Increasing the order of singular functions further improves accuracy, with errors for all four variations of the geometry under $2\%$ using third-order singular functions. The number of unknowns for each geometry at each order is provided in Table \ref{table:GCPW-unknowns}, along with the number of tetrahedra.

\begin{figure}[t!]
    \centering
    \includegraphics[width=0.77\linewidth]{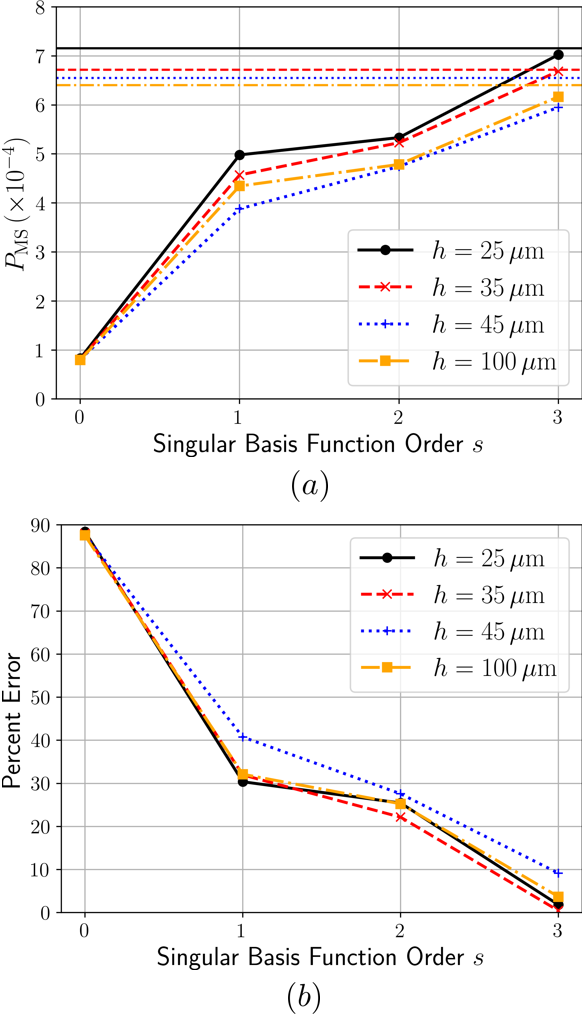}
    \caption{Metal-substrate participation ratio $P_\mathrm{MS}$ for the GCPW geometry. The ratio is calculated over only the $100 \, \mu \mathrm{m}$ central region highlighted blue in Fig. \ref{fig:gcpw}. As in Fig. \ref{fig:GCPW-cap}, $p=2$ is used in all cases, and $s=0$ represents no singular functions being used. As in Fig. \ref{fig:GCPW-cap}, percent errors between numerical and analytical results (represented by horizontal lines) are provided in (b).}
    \label{fig:GCPW-sp}
\end{figure}

In addition to the PUL capacitance, we extract participation ratios for the GCPW and validate against analytical solutions. As mentioned in Section \ref{sec:intro}, participation ratios are one of the key quantities used in predicting superconducting qubit coherence times through electromagnetic simulations. The participation ratio describes the proportion of the electrical energy found within a thin interface layer between materials. We focus on the metal-substrate (MS) interface, for which the participation ratio is given by
\begin{align}
    P_\mathrm{MS} = \frac{\int_{\Omega_\mathrm{i}} \epsilon |\mathbf{E}|^2 \, \mathrm{d}V}{\int_{\Omega} \epsilon |\mathbf{E}|^2 \, \mathrm{d}V},
    \label{eq:part-ratio-vol}
\end{align}
where $\Omega$ is the volume of the entire simulation region, and $\Omega_i$ is the volume of the interface layer. The interface layer used in \cite{2026_Wang_SIE} is only $3 \, \mathrm{nm}$ thick, so including it in the simulation region would require meshing it with extremely small tetrahedra, negating the benefit of the singular basis functions and vastly increasing computational cost. Instead, we integrate over the metal surface containing the thin layer, so
\begin{align}
    P_\mathrm{MS} = \frac{ t_\mathrm{i} \int_{\Gamma_\mathrm{i}} \epsilon |\mathbf{E}_\mathrm{i}|^2 \, \mathrm{d}S}{\int_{\Omega} \epsilon |\mathbf{E}|^2 \, \mathrm{d}V},
    \label{eq:part-ratio}
\end{align}
where $\Gamma_\mathrm{i}$ is the surface where the interface would lie, and $t_\mathrm{i}$ is the thickness of the layer. Further, $\mathbf{E}_\mathrm{i}$ is the electric field within the interface layer, which is rescaled to account for differences in the material properties relative to the substrate. In particular, if $\epsilon_\mathrm{s}$ is the relative permittivity of the substrate and $\epsilon_\mathrm{i}$ is that of the interface layer, we have \cite{2017_Gambetta_Surface_Loss}
\begin{align}
    \mathbf{E}_\mathrm{i} = \frac{\epsilon_\mathrm{s}}{\epsilon_\mathrm{i}} (\mathbf{E} \cdot \mathbf{\hat{n}} ) \mathbf{\hat{n}}  - \mathbf{\hat{n}} \times \big( \mathbf{\hat{n}} \times \mathbf{E} \big),
    \label{eq:field-rescale}
\end{align}
where $\mathbf{\hat{n}}$ is the unit vector normal to the metal surface. In \cite{2026_Wang_SIE}, $P_\mathrm{MS}$ is only calculated along a $100 \, \mu \mathrm{m}$ region of the geometry, which is highlighted in blue in Fig. \ref{fig:gcpw}. Therefore, both the surface and volume integrals in (\ref{eq:part-ratio}) are consolidated to this region to maintain consistency with those results.

In Fig. \ref{fig:GCPW-sp}(a), we compare the participation ratios extracted from our numerical simulations against analytical results from \cite{2026_Wang_SIE}. These ratios are calculated for an interface layer with thickness $t_\mathrm{i} = 3 \, \mathrm{nm}$ and relative permittivity $\epsilon_\mathrm{i} = \epsilon_\mathrm{s} =  11.9$, as done in \cite{2026_Wang_SIE}. When $(p,s)= (2,0)$ and no singular functions are used, the resulting $P_\mathrm{MS}$ values are extremely far from the analytical solutions, with the errors in Fig. \ref{fig:GCPW-sp}(b) at almost $90\%$ in all four cases. This indicates that $P_\mathrm{MS}$ is dominated by the singularities, which are represented poorly by the standard basis functions alone. Introducing the first-order singular functions with $s=1$ increases $P_\mathrm{MS}$ by a factor of four or more across the different substrate heights, bringing it much closer to the analytical solution. Further improvements are observed by increasing the order of singular functions, with errors under $10\%$ for all four heights when $s=3$, and under $4 \%$ for three of the four. 

We note that this disagreement may be partially due to the conducting box, which is not included in the analytical solution. Further, experimental efforts to characterize losses due to interface layers in CPW resonators typically see discrepancies of $>30\%$ between simulations and measurements \cite{2018_Caulsine_CPW_Interfaces, 2023_Crowley_CPW_Losses}. Our results fall well within those error margins, meaning the singular basis functions provide more than sufficient accuracy to support the design of these devices.

\begin{table}[t]
\caption{Number of unknowns for the GCPW at each substrate height $h$ and singular function order $s$} 
\renewcommand{\arraystretch}{1.5}
\centering
\newcolumntype{C}{>{\centering\arraybackslash} m{0.04\textwidth}}
\newcolumntype{D}{>{\centering\arraybackslash} m{0.075\textwidth}}
\newcolumntype{M}{>{\centering\arraybackslash} m{0.085\textwidth}}
\footnotesize
\label{table:GCPW-unknowns}
\begin{tabular}{C|D|D|D|M}
 \hline
 \hline
  & $h = 25 \, \mu\mathrm{m}$ $6583$ tets& $h = 35 \, \mu\mathrm{m}$ $7422$ tets & $h = 45 \, \mu\mathrm{m}$ $7966$ tets & $h = 100 \, \mu\mathrm{m}$ $9274$ tets\\
 \hline
 \hline
 $s=0$ & $35,242$ & $39,966$ & $43,530$ & $51,606$\\
 \hline
 $s=1$ & $40,916$ & $45,856$ & $49,218$ & $57,684$\\
 \hline
 $s=2$ & $57,366$ & $62,914$ & $65,692$ & $75,292$\\
 \hline
 $s=3$ & $88,568$ & $95,254$ & $96,908$ & $108,704$ \\
 \hline
 \hline
\end{tabular}
\end{table}

\subsection{Numerical Validation}
\label{subsec:val-num}

\begin{figure}
    \centering
    \includegraphics[width=0.9\linewidth]{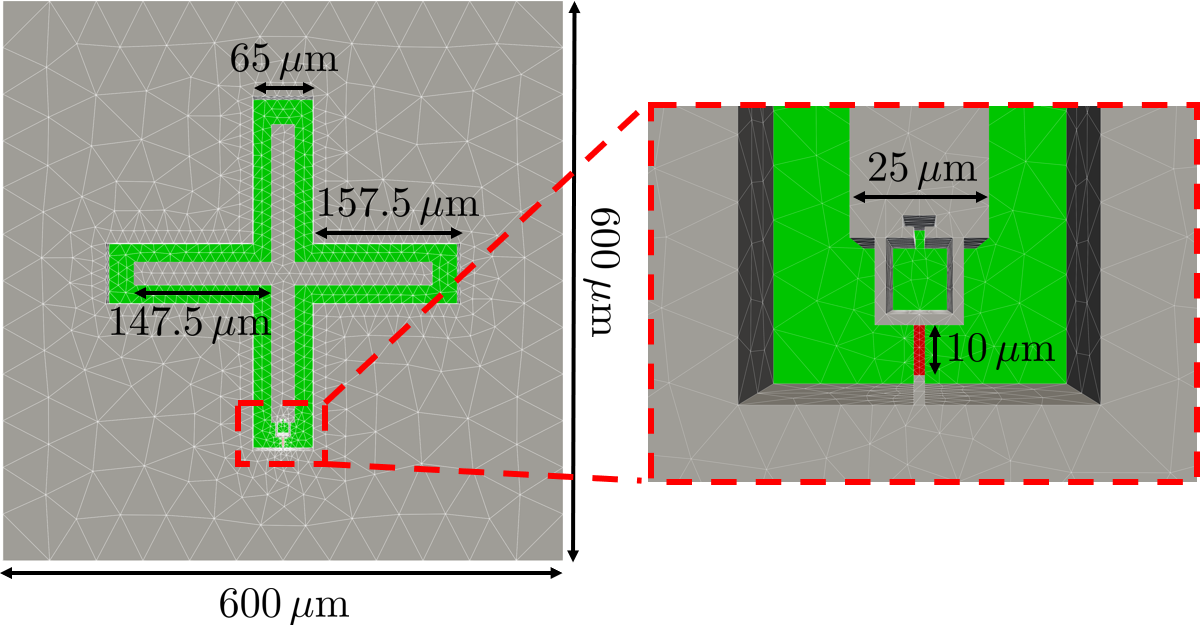}
    \caption{The Xmon geometry with the mesh shown in white. The conductors are colored gray, while the exposed substrate ($\epsilon_r = 11.45$) is in green. The inset shows the region around the superconducting quantum interference device (SQUID) loop, which typically contains two Josephson junctions. For simplicity, we omit these and instead drive the device from a lumped port, which is highlighted red. The metal layer is $30 \, \mu \mathrm{m}$ tall. The geometry is enclosed by a perfectly conducting box, which extends $70 \, \mu \mathrm{m}$ above the top of the metal layer and to the bottom of the substrate, which is $125 \, \mu \mathrm{m}$ thick.}
    \label{fig:xmon}
\end{figure}

\begin{figure*}
    \centering
    \includegraphics[width=0.95\linewidth]{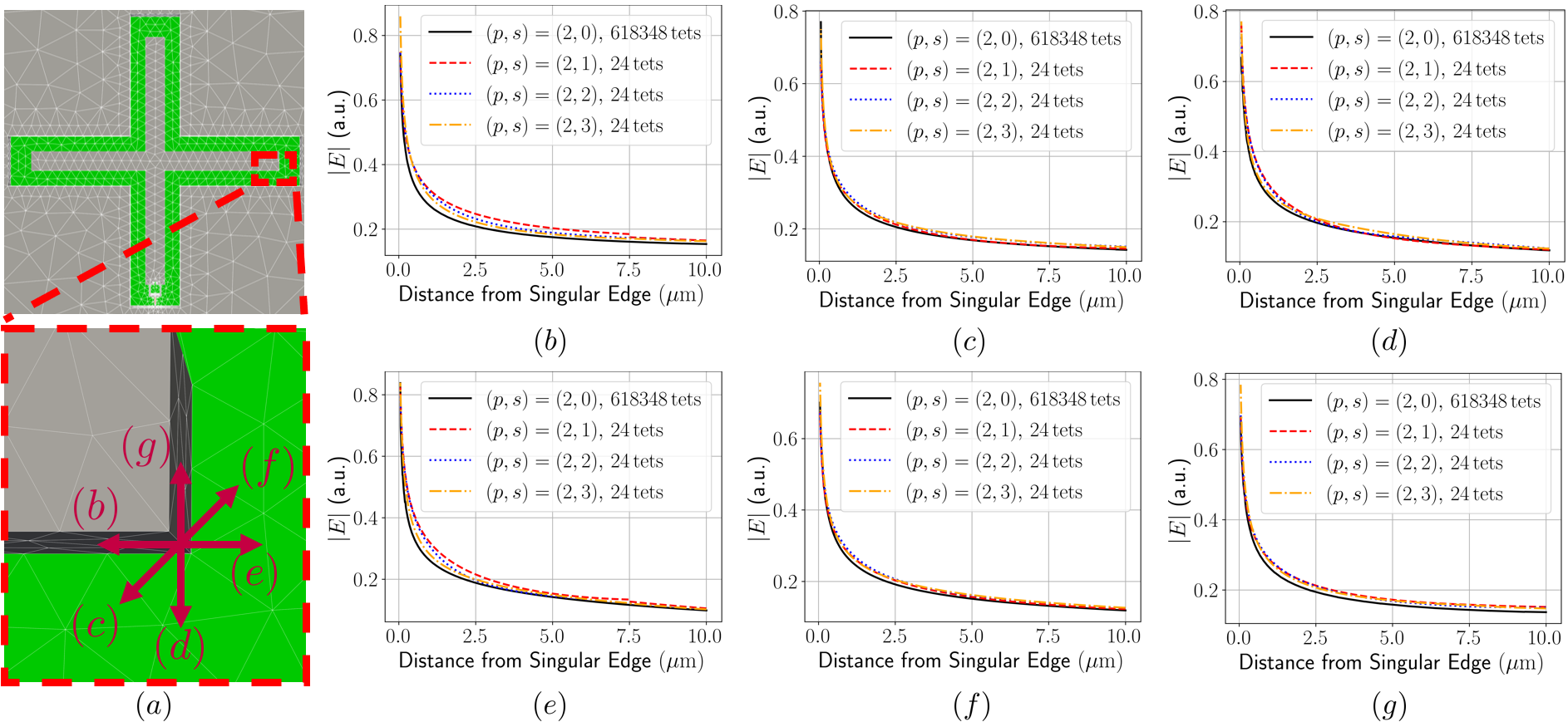}
    \caption{(a) The Xmon geometry with measurement lines protruding from the center of an edge normal to the plane of the geometry. The set of arrows indicate the directions of the measurement lines used in (b) to (g) where the magnitude of the electric field is plotted. The solid black lines come from a simulation with no singular basis functions that uses an extremely fine mesh near the center of the singular edge. The other lines are from simulations using various orders of singular functions. The number of tetrahedra in an equivalent region near the singularity are provided in the legend.}
    \label{fig:xmon-zLine}
\end{figure*}

In addition to these analytical results, we validate our approach against standard FEM solutions using extremely fine meshing in the vicinity of a singular feature. These tests are performed on the Xmon qubit geometry shown in Fig. \ref{fig:xmon}, which is adapted from the design of \cite{2024_Ma_Xmon}. These devices typically have extremely thin $\mathcal{O}(100 \, \mathrm{nm})$ metal layers, leading to contributions from multiple singular features in this region and complicating the field structure. To simplify this behavior and obtain better consistency with the analytical results in (\ref{eq:sing-E}) to (\ref{eq:sing-H-approx}) for the purposes of validation, we increase the height of the metal layer to $30 \, \mu \mathrm{m}$. 

Because the metal layer has finite thickness, there is a mix of metal edges at $\phi^* = 90 \degree$ and $\phi^* = 270 \degree$ angles throughout the geometry, where $\phi^*$ is as defined in Fig. \ref{fig:singEdge}. To match the expected behavior, the $270 \degree$ edges are treated as singular with $\nu = 2/3$, while the $90 \degree$ angles are nonsingular, and thus only require the standard basis functions to be used.

As shown by Fig. \ref{fig:xmon-zLine}(a), we focus on field behavior originating from a single point along a singular feature. In particular, we save the electric field at $1000$ equidistant points along lines extending in the directions shown in Fig. \ref{fig:xmon-zLine}(a), where the first point on each line is positioned $50 \, \mathrm{nm}$ away from the feature. The field magnitudes at each of these points is plotted in Figs. \ref{fig:xmon-zLine}(b) to \ref{fig:xmon-zLine}(g). 

To validate the behavior of the singular basis functions along these lines, we perform simulations using our FEM implementation with only the standard basis functions ($s=0$) on an extremely refined mesh in a cylindrical region surrounding the feature. We start by defining a cylindrical meshing region with radius $2 \,\mathrm{nm}$ and height $4 \,\mathrm{nm}$ centered about the midpoint of the edge, and meshing it with tetrahedra with maximum edge lengths of $0.25 \, \mathrm{nm}$. This process is repeated with the cylinder radius, height, and maximum edge length doubled at each iteration until the radius exceeds $10 \, \mu \mathrm{m}$. In total, these cylinders contain $618,348$ tetrahedra, as indicated in the legend of Figs. \ref{fig:xmon-zLine}(b) to \ref{fig:xmon-zLine}(g).

When the singular basis functions are included, we use the same mesh density in this region as in the rest of the geometry. However, we define an outer cylindrical region to allow us to count the number of tetrahedra near the singularity, which is only $24$. In Figs. \ref{fig:xmon-zLine}(b) to \ref{fig:xmon-zLine}(g), we compare the electric field magnitudes in the dense mesh with $(p,s) = (2,0)$ to the behavior produced by the singular functions in the coarse mesh with varying $s$. In general we find that the singular basis functions capture the underlying behavior of the dense mesh with several orders of magnitude fewer elements needed. On some of the lines where the first-order singular basis functions are not as accurate, such as Fig. \ref{fig:xmon-zLine}(b) and \ref{fig:xmon-zLine}(e), increasing $s$ leads to significantly better behavior. Along some of the other lines, the first-order singular basis functions capture the underlying behavior quite accurately, and increasing $s$ does not produce significant changes in the resulting field magnitudes.

\begin{figure*}
    \centering
    \includegraphics[width=0.95\linewidth]{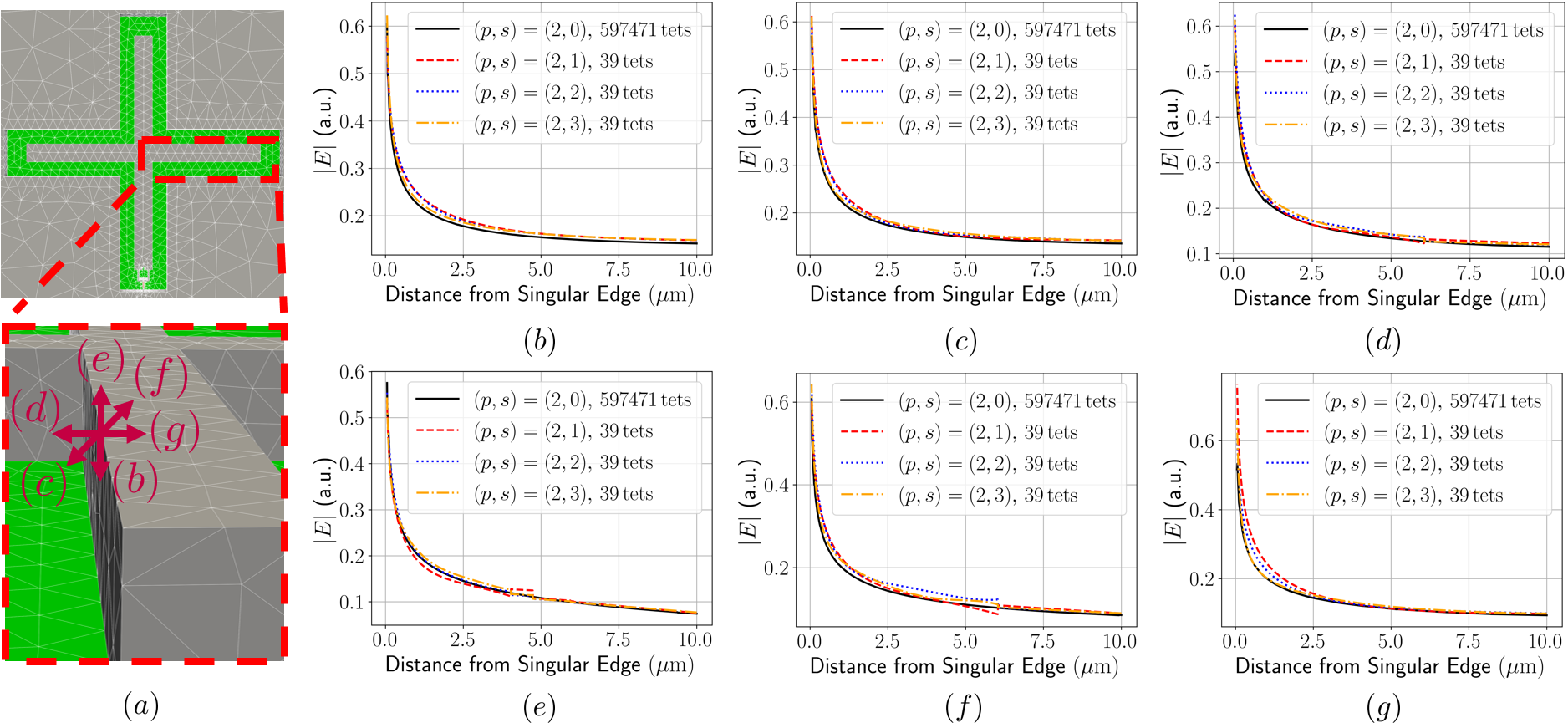}
    \caption{The Xmon geometry with measurement lines protruding from the center of an edge tangential to the plane of the geometry. As in Fig. \ref{fig:xmon-zLine}, the electric field magnitude is plotted in (b) to (g) at points along lines marked in the inset of (a). Results are shown for the standard basis functions on an extremely fine mesh in the surrounding region, and for the singular functions with no refinement in the mesh around this region.}
    \label{fig:xmon-xLine}
\end{figure*}

We repeat this process for a second set of lines along another edge in Fig. \ref{fig:xmon-xLine}, where the edge is along the top surface of the metal layer in this case. The same cylindrical meshing procedure is used here, where $597,471$ tetrahedra are contained within this region for the highly refined mesh. For the mesh used with the singular functions, only $39$ tetrahedra are present within this region. As in Fig. \ref{fig:xmon-zLine}, the results in Figs. \ref{fig:xmon-xLine}(b) to \ref{fig:xmon-xLine}(g) indicate that the singular basis functions can represent this behavior accurately at a much lower computational cost. 

\subsection{Participation Ratio Extraction}
\label{subsec:part-ratios}

\begin{figure}[t!]
    \centering
    \includegraphics[width=0.9\linewidth]{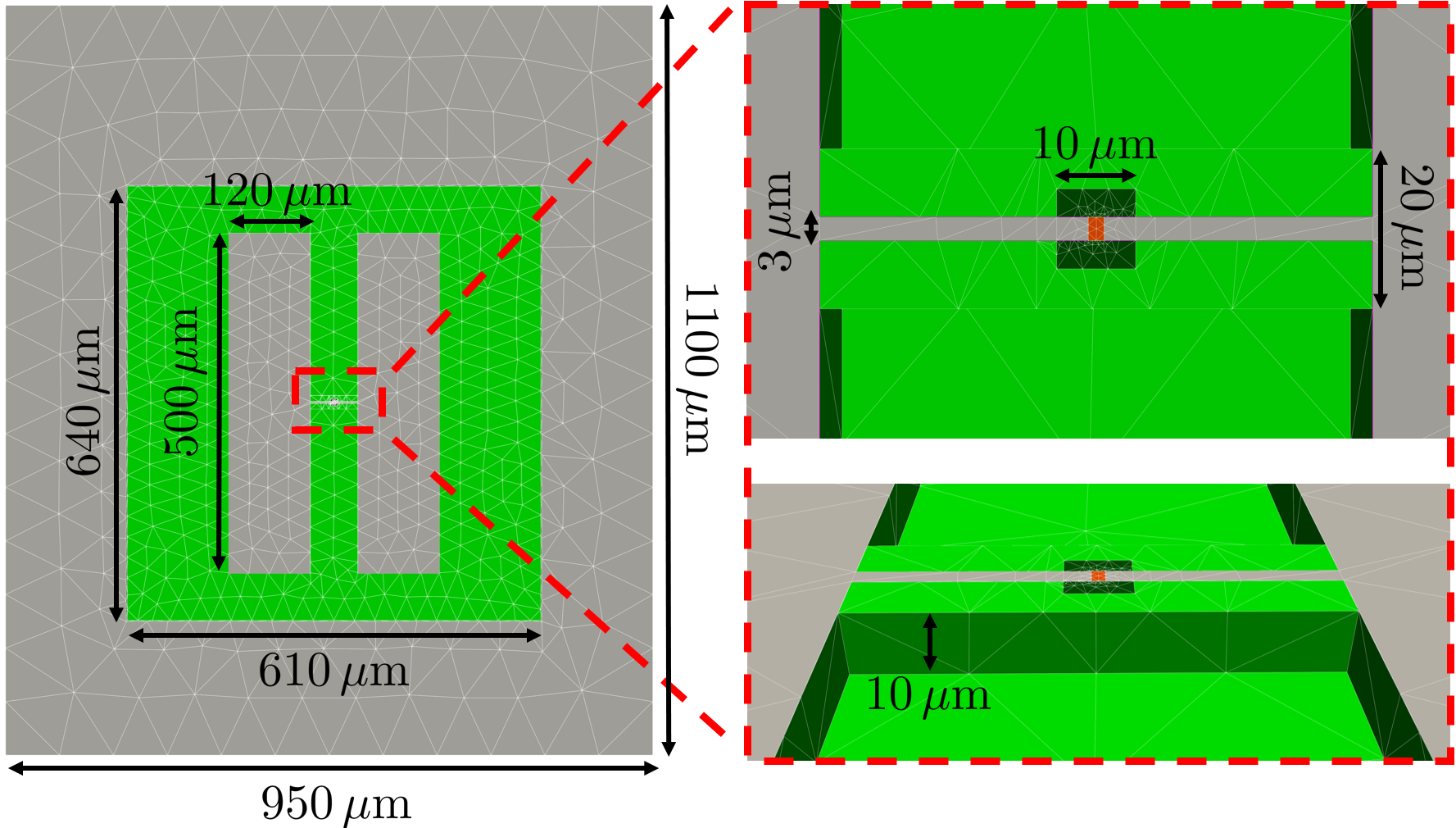}
    \caption{The qubit geometry with the mesh overlaid. The geometry consists of a central junction region (shown in the inset) connected by $3 \, \mu \mathrm{m}$ wide leads to two pad. The conductors (colored gray) are all $100 \, \mathrm{nm}$ thick and sit on a $300 \, \mu \mathrm{m}$ substrate with $\epsilon_r = 11.45$. Outside of a $20 \, \mu \mathrm{m}$ wide region between the ports, the exposed substrate (colored green) is trenched a depth of $10 \, \mu \mathrm{m}$. The geometry is enclosed by a perfectly conducting box flush with the outer bounds of the ground plane, which extends $300 \, \mu \mathrm{m}$ above the conductors and to the bottom of the substrate. Because the geometry is driven using a lumped port (colored red) in place of the Josephson junction, we eliminate contributions to participation ratios from the $10 \, \mu \mathrm{m}$ square region surrounding the lumped port (colored dark green).}
    \label{fig:qubitGeom}
\end{figure}

To demonstrate the utility of these functions for designing superconducting qubits, we use them to extract participation ratios for the qubit shown in Fig. \ref{fig:qubitGeom}, which is adapted from \cite{2017_Gambetta_Surface_Loss}. As shown in the inset of Fig. \ref{fig:qubitGeom}, the geometry has a $10 \, \mu \mathrm{m}$ trench where the exposed substrate has been lowered. Trenching is a critical technique in superconducting qubit design for mitigating losses in the interface layers by shifting the layers away from singular features \cite{2011_Martinis_Participation, 2017_Gambetta_Surface_Loss, 2019_Woods_Interface_Loss, 2018_Caulsine_CPW_Interfaces}. While the surface integral equation solver developed in \cite{2026_Wang_SIE} could rapidly evaluate participation ratios for fully planar geometries such as the GCPW of Fig. \ref{fig:gcpw}, it cannot be applied to a trenched geometry like this one. In contrast, accounting for the trench in our approach requires no modifications.

As in Section \ref{subsec:val-ana}, we calculate participation ratios using the surface-based approach of (\ref{eq:part-ratio}). In addition to the MS interface, we calculate these ratios for the substrate-air (SA) and metal-air (MA) interfaces, where we use $t_\mathrm{i} = 3 \, \mathrm{nm}$ and $\epsilon_\mathrm{i} = 5$ in all cases as in \cite{2017_Gambetta_Surface_Loss}. As in that work, we model the substrate with $\epsilon_\mathrm{s} = 11.45$. Due to the difference in material properties between the substrate, air, and the interface layers, the electric field at interface surfaces must be rescaled using (\ref{eq:field-rescale}) when calculating participation ratios. The contributions from the ground plane are relatively small, so we omit them and only account for the MS and MA interfaces on the pads and their respective leads. Further, we discard the contributions from $100 \, \mathrm{nm}$ tall surfaces along the side of the metal layer due to difficulty in extracting them using commercial tools. Additionally, we eliminate contributions from a $10 \, \mu \mathrm{m}$ square region surrounding the port surface. This port is an artificial surface used to drive the geometry, meaning that this region is not reflective of the physical geometry and thus its contribution is not relevant to the qubit design process.

We calculate these ratios using the mesh pictured in Fig. \ref{fig:qubitGeom}, which contains $23,052$ tetrahedra. As in Section \ref{subsec:val-num}, we use singular basis functions with $\nu = 2/3$ at nodes and edges along $\phi^* = 270 \degree$ angles in the conductive volumes, while only the standard basis functions are used elsewhere. We use $(p, s) = (2,3)$, leading to a discretized system with $2.940 \times 10^5$ unknowns. The ratios are then extracted from a $5 \, \mathrm{GHz}$ simulation driven from the lumped port, which is treated as an open-circuited current source. 

To validate our results, we simulate the same geometry in HFSS with an extremely fine mesh. In the HFSS simulation  we include the interface layers as volumes in the geometry, but increase their thickness to $30 \, \mathrm{nm}$. This modification is made to simplify the meshing process, and the participation ratios calculated using (\ref{eq:part-ratio-vol}) are divided by $10$ to account for this difference in thickness. 

\begin{table}[t]
\caption{Participation ratios and resulting quality factor for the qubit geometry of Fig. \ref{fig:qubitGeom}} 
\renewcommand{\arraystretch}{1.5}
\centering
\newcolumntype{D}{>{\centering\arraybackslash} m{0.09\textwidth}}
\newcolumntype{M}{>{\centering\arraybackslash} m{0.105\textwidth}}
\newcolumntype{N}{>{\centering\arraybackslash} m{0.068\textwidth}}
\footnotesize
\label{table:qubit-part}
\begin{tabular}{M||D|D|N}
 \hline
 \hline
  &Ansys HFSS& Singular Functions & Percent Difference\\
 \hline
 \hline
 Left pad $P_\mathrm{MS}$ & $6.553 \times 10^{-5}$ & $6.829 \times 10^{-5}$ & $4.12\%$ \\
 \hline
 Right pad $P_\mathrm{MS}$ & $6.550 \times 10^{-5}$ & $6.830 \times 10^{-5}$ & $4.19\%$ \\
 \hline
 Left pad $P_\mathrm{MA}$ & $1.086 \times 10^{-6}$ & $8.117 \times 10^{-7}$ & $-28.91\%$ \\
 \hline
 Right pad $P_\mathrm{MA}$ & $1.091 \times 10^{-6}$ & $8.054 \times 10^{-7}$ & $-30.12\%$\\
 \hline
 $P_\mathrm{SA}$ & $4.238 \times 10^{-5}$ & $5.231 \times 10^{-5}$ & $20.97\%$\\
 \hline
 \hline
 \textbf{Quality factor} & $\mathbf{5.695 \times 10^6}$ & $\mathbf{5.249 \times 10^6}$ & $\mathbf{-8.16 \%}$ \\
 \hline
 \hline
\end{tabular}
\end{table}

Typically, HFSS simulations are initialized with an extremely coarse mesh, and an iterative adaptive refinement procedure is used to make targeted improvements to the mesh in areas where error is large. However, this algorithm is not well-suited to focusing refinements towards improving accuracy in complicated objective functions such as participation ratios. Because of this, we instead seed our initial mesh so that the interface layers contain tetrahedra with maximum edge lengths of $1 \, \mu \mathrm{m}$. This produces an initial mesh with $1.258 \times 10^7$ tetrahedra, which is then used as the input to the adaptive refinement process. We allow this process to iterate until the difference between all of the participation ratios in successive passes is below $1 \%$, which occurs on the fourth iteration. This final mesh contains $2.539 \times 10^7$ tetrahedra, which produces a system with $1.571 \times 10^8$ unknowns using basis functions of order $p=2$ in HFSS (note that HFSS uses different terminology than \cite{2015_Jin_FEM} and this work, so these are referred to as ``first-order'' basis functions in HFSS). 

Despite the far coarser mesh, the singular basis functions allow us to achieve relatively strong agreement with HFSS, as captured in Table \ref{table:qubit-part}. Most of the energy is found in the MS interfaces, for which we see very good agreement. While the differences are larger for the SA and MA interfaces, these values are still within the level of accuracy desired by engineers in this field to improve qubit designs. As discussed in Section \ref{subsec:val-ana}, experimental measurements of quality factors for CPW resonators typically vary relative to the predicted values from the participation ratios by over $30 \%$ \cite{2018_Caulsine_CPW_Interfaces, 2023_Crowley_CPW_Losses}, and similar results have been observed for qubits \cite{2025_Bland_2D_ms_coherence}. Therefore, the values predicted by both approaches show good enough agreement to support the qubit design process. To emphasize this, we also include the expected quality factor $Q$ calculated from the participation ratios predicted by both methods in Table \ref{table:qubit-part}. This calculation is done assuming all interface layers have $\tan \delta = 10^{-3}$, where $1/Q = \sum_i P_i \tan \delta_i$ for each interface layer $i$ \cite{2018_Caulsine_CPW_Interfaces}. The difference between these quality factors is only $8.16 \%$, meaning the simulations show very strong agreement compared to the typical variations between simulated and experimental results. We note that this percent difference (and all those in Table \ref{table:qubit-part}) are calculated with respect to the average of the results from the two methods.

We provide a head-to-head comparison between the timing and memory constraints of our singular basis functions against HFSS in Table \ref{table:cost}. Due to memory constraints, the HFSS simulations were ran on the Google Cloud Platform using h4d-highmem-192 machines where $800$ cores and $13.5 \, \mathrm{TB}$ of total memory were available for the simulation, while the simulations using our singular basis functions were ran using $16$ cores and $32 \, \mathrm{GB}$ of total memory on the Negishi cluster provided by the Rosen Center for Advanced Computing at Purdue University. The timing data for the HFSS results is summed over all four adaptive passes, while the memory usage and mesh information is listed for only the final pass. As shown in the table, the HFSS solution process required $21.27 \, \mathrm{h}$  to complete, while our simulation concluded in $196 \, \mathrm{s}$. Accounting for the disparity in number of cores used, this represents over a 19,500-fold improvement in CPU-hours for this application. By providing the desired level of accuracy at a far lower computational cost, this approach can potentially enable optimization of qubit designs to achieve improved coherence times, leading to better performing quantum computers. 

\begin{table}[t]
\caption{Comparison of Participation Ratio Simulation Costs} 
\renewcommand{\arraystretch}{1.5}
\centering
\newcolumntype{D}{>{\centering\arraybackslash} m{0.095\textwidth}}
\newcolumntype{M}{>{\centering\arraybackslash} m{0.16\textwidth}}
\footnotesize
\label{table:cost}
\begin{tabular}{M||D|D}
 \hline
 \hline
  &Ansys HFSS& Singular Functions\\
 \hline
 \hline
 Number of tetrahedra & $2.539 \times 10^{7}$ & $2.305 \times 10^{4}$ \\
 \hline
 Number of unknowns & $1.571 \times 10^{8}$ & $2.940 \times 10^{5}$ \\
 \hline
 Total simulation time (s) & $7.656 \times 10^4$ & $196.02$  \\
 \hline
 Total simulation time (CPU-hours) & $1.701 \times 10^4$ & $0.871$ \\
 \hline
 Peak memory usage & $6.23 \, \mathrm{TB}$ & $27.24 \, \mathrm{GB}$ \\
 \hline
 \hline
\end{tabular}
\end{table}

\section{Conclusion}
\label{sec:conclusion}
This work developed sets of basis functions for representing singular behavior in complex geometries containing many singular features. These functions were constructed by leveraging ideas developed for singular basis functions in triangles, and additional novel functions were created to model behavior emerging from tetrahedron edges along a singular feature. Higher-order variations of these functions were also derived to model singular behavior with improved accuracy. Analytical solutions for a simple geometry were used to demonstrate the substantial accuracy improvements from incorporating these functions. Additional validation was performed against traditional FEM on a highly refined mesh, where the field behavior produced by the singular basis functions showed strong agreement even when the number of tetrahedra was reduced by several orders of magnitude. Further, a potential application of the method to designing superconducting qubits was shown by demonstrating the computational benefits of using these functions to extract participation ratios.

In the future, the singular basis functions developed in this work can be combined with optimization techniques to better design qubits that are predicted to exhibit improved qubit coherence times. The utility of these singular basis functions can be improved by supplementing them with functions for other types of volumetric elements, such as triangular prisms. These could be particularly useful for maintaining high-quality mesh elements for applications that involve thin structures, such as for superconducting qubits. Further, adaptive mesh refinement strategies can be developed to work with these singular basis functions to obtain further improvements in accuracy beyond just relying on increasing the order of the singular basis functions. Additional accuracy improvements can be obtained by developing functions for modeling behavior near 3D corners. Another area of future work would be to expand the set of singular basis functions to also enable modeling of singular magnetic fields, which can be particularly important in other electromagnetic scattering applications. 

\appendices
\section{}
\label{app:edge-sing}
In this appendix, we start from the edge-singular gradient functions of (\ref{eq:sing-edge-grad}) and derive the tangential component of the functions along their associated faces, which is given by (\ref{eq:sing-edge-grad-tan}). We begin by expanding (\ref{eq:sing-edge-grad}) in the form
\begin{multline}
    \mathbf{N}^{ijk,g_1} = (\nu - 1) (1-\xi_i - \xi_j)^{\nu-2} \xi_i \xi_j \xi_k \big( \nabla \xi_i + \nabla\xi_j \big) \\ + \big( 1 - (1-\xi_i - \xi_j)^{\nu-1} \big) \bigg[ \xi_i \xi_j \nabla \xi_k + \xi_i \xi_k \nabla \xi_j + \xi_j \xi_k \nabla \xi_i \bigg].
    \label{eq:sing-edge-grad-expand}
\end{multline}
On the face, $\xi_k = 1 - \xi_i - \xi_j$, so we can write
\begin{multline}
    \mathbf{N}^{ijk,g_1} \bigg|_{\mathrm{face} \, ijk} = (\nu - 1) (1-\xi_i - \xi_j)^{\nu-1} \xi_i \xi_j \big( \nabla \xi_i + \nabla\xi_j \big) \\ + \big( 1 - (1-\xi_i - \xi_j)^{\nu-1} \big) \bigg[ \xi_i \xi_j \nabla \xi_k + \xi_i \xi_k \nabla \xi_j + \xi_j \xi_k \nabla \xi_i \bigg].
\end{multline}

We are specifically interested in the behavior of the function tangential to the face. As established in Section \ref{subsubsec:node-funcs}, the unit vector normal to the face is $\mathbf{\hat{n}} = \nabla \xi_\ell / | \nabla \xi_\ell |$, where node $\ell$ is opposite the face. To simplify the subsequent expressions, we introduce the notation $\nabla_t \xi_i \equiv  \mathbf{\hat{n}} \times \mathbf{\hat{n}} \times \nabla \xi_i$ to describe the portion of $\nabla \xi_i$ tangential to face $ijk$. Then due to the dependency relation between the barycentric coordinates, we have $\nabla_t \xi_i + \nabla_t \xi_j + \nabla_t \xi_k = 0$, leading to
\begin{multline}
    \mathbf{\hat{n}} \times \mathbf{\hat{n}} \times\mathbf{N}^{ijk,g_1} \bigg|_{\mathrm{face} \, ijk} = (1 - \nu) (1-\xi_i - \xi_j)^{\nu-1} \xi_i \xi_j  \nabla_t \xi_k \\ + \big(1 - (1-\xi_i - \xi_j)^{\nu-1} \big) \bigg[ \xi_i \xi_j \nabla_t \xi_k + \xi_i \xi_k \nabla_t \xi_j + \xi_j \xi_k \nabla_t \xi_i \bigg],
\end{multline}
or equivalently,
\begin{multline}
    \mathbf{\hat{n}} \times \mathbf{\hat{n}} \times\mathbf{N}^{ijk,g_1} \bigg|_{\mathrm{face} \, ijk} =  \big( 1 - \nu  (1-\xi_i - \xi_j)^{\nu-1} \big) \xi_i \xi_j  \nabla_t \xi_k \\ + \big( 1 - (1-\xi_i - \xi_j)^{\nu-1} \big) \bigg[ \xi_i \xi_k \nabla_t \xi_j + \xi_j \xi_k \nabla_t \xi_i \bigg].
\end{multline}
If we apply $\xi_k = 1 - \xi_i - \xi_j$ once again, we obtain
\begin{multline}
    \mathbf{\hat{n}} \times \mathbf{\hat{n}} \times\mathbf{N}^{ijk,g_1} \bigg|_{\mathrm{face} \, ijk} = \big(1 - \nu  (1-\xi_i - \xi_j)^{\nu-1} \big) \xi_i \xi_j  \nabla_t \xi_k \\ + (\xi_k - \xi_k^{\nu}) \bigg[ \xi_i \nabla_t \xi_j + \xi_j \nabla_t \xi_i \bigg].
    \label{eq:sing-edge-grad-expand-tan}
\end{multline}

The behavior of (\ref{eq:sing-edge-grad-expand-tan}) can be described in terms of three terms. The first is $\xi_i \xi_j \nabla_t \xi_k$, which is constant up to the weighting factor $\xi_i \xi_j$, and normal to the edge. As discussed in Section \ref{subsubsec:edge-funcs} of the main text, this cannot be eliminated in combination with the standard basis functions alone due to the weighting factor. However, this issue can be resolved by simply using standard higher-order basis functions.

The remaining two terms are the singular term normal to the edge, and the $\xi_k - \xi_k^\nu$ term tangential to the edge. Close to the edge, $\xi_k = 1 - \xi_i - \xi_j \rightarrow 0$, and the singular term dominates, leading to the result in (\ref{eq:sing-edge-grad-tan}). Further from the edge, the tangential term still remains relatively small for typical values of $\nu$. For example, if the singular feature is a conductive wedge with $\phi^* = 3 \pi/2$, $\nu = 2/3$ then the maximum value of $|\xi_k - \xi_k^v|$ is $0.148$ at $\xi_k = 0.296$. Because the contribution from this term is typically minimal, the form of (\ref{eq:sing-edge-grad-tan}) should generally be a good approximation to the tangential portion of (\ref{eq:sing-edge-grad}) along the face.

\section{}
\label{app:edge-sing-rot}
Here, we show how the curl of the edge-singular rotational function in (\ref{eq:sing-edge-rot}) can be approximated by (\ref{eq:sing-edge-rot-curl-approx}). We start by expanding the curl in the form
\begin{multline}
    \nabla \times \mathbf{N}^{ijkl,r_1} = \big( (1-\xi_i - \xi_j)^\nu - 1 \big) \xi_i \xi_j \nabla \times \mathbf{N}^{k\ell} \\ + \nabla \bigg( \big( (1-\xi_i - \xi_j)^\nu - 1 \big) \xi_i \xi_j \bigg)\times \mathbf{N}^{k\ell},
\end{multline}
which is equivalent to
\begin{multline}
    \nabla \times \mathbf{N}^{ijkl,r_1} = \big( (1-\xi_i - \xi_j)^\nu - 1 \big) \xi_i \xi_j \nabla \times \mathbf{N}^{k\ell} \\ + \bigg( \big( (1-\xi_i - \xi_j)^\nu - 1 \big) \big( \xi_i \nabla \xi_j + \xi_j \nabla \xi_i \big) \bigg)\times \mathbf{N}^{k\ell} \\ - \nu (1-\xi_i - \xi_j)^{\nu-1} \xi_i \xi_j (\nabla\xi_i + \nabla \xi_j) \times \mathbf{N}^{k\ell}.
    \label{eq:sing-edge-rot-curl-expand}
\end{multline}
To simplify the following expressions, it is convenient to use the shorthand $\mathbf{e}_i = \nabla \xi_j \times \nabla \xi_k, \, \mathbf{e}_j = \nabla \xi_k \times \nabla \xi_i, \, \mathbf{e}_k = \nabla \xi_i \times \nabla \xi_j $ defined in Section \ref{subsec:implementation}. Using the dependency relationship to eliminate $\nabla \xi_\ell$ terms, this allows us to write
\begin{align}
    \nabla \times \mathbf{N}^{k\ell} = -2 \big( \mathbf{e}_i - \mathbf{e}_j),
\end{align}
\begin{align}
    \nabla \xi_i \times \mathbf{N}^{k \ell} = -(1 - \xi_i - \xi_j) \mathbf{e}_j + \xi_k \mathbf{e}_k,
\end{align}
\begin{align}
    \nabla \xi_j \times \mathbf{N}^{k \ell} = (1 - \xi_i - \xi_j) \mathbf{e}_i - \xi_k \mathbf{e}_k.
\end{align}
Applying these definitions to (\ref{eq:sing-edge-rot-curl-expand}) leads to
\begin{multline}
    \nabla \times \mathbf{N}^{ijkl,r_1} = -2 \big( (1-\xi_i - \xi_j)^\nu - 1 \big) \xi_i \xi_j \big(\mathbf{e}_i - \mathbf{e}_j ) \\ 
    + \big( (1-\xi_i - \xi_j)^\nu - 1 \big)(1-\xi_i - \xi_j) \big(\xi_i \mathbf{e}_i - \xi_j \mathbf{e}_j ) \\
    + \big( (1-\xi_i - \xi_j)^\nu - 1 \big) \xi_k (\xi_j - \xi_i) \mathbf{e}_k \\
    - \nu \big( 1 - \xi_i - \xi_j \big)^{\nu} \xi_i \xi_j (\mathbf{e}_i - \mathbf{e}_j ),
\end{multline}
where the first and fourth lines can be combined to obtain
\begin{multline}
    \nabla \times \mathbf{N}^{ijkl,r_1} = \big( 2 - (2 + \nu) (1-\xi_i - \xi_j)^\nu \big) \xi_i \xi_j \big(\mathbf{e}_i - \mathbf{e}_j ) \\ 
    + \big( (1-\xi_i - \xi_j)^\nu - 1 \big)(1-\xi_i - \xi_j) \big(\xi_i \mathbf{e}_i - \xi_j \mathbf{e}_j ) \\
    + \big( (1-\xi_i - \xi_j)^\nu - 1 \big) \xi_k (\xi_j - \xi_i) \mathbf{e}_k.
    \label{eq:sing-edge-rot-curl-full}
\end{multline}

In this form, it is difficult to gain insight into the structure of the function. To simplify things further, we analyze the behavior along the faces touching singular edge $ij$. On face $ijk$, using $\xi_k = 1 - \xi_i - \xi_j$ gives
\begin{multline}
    \nabla \times \mathbf{N}^{ijkl,r_1} \bigg |_\mathrm{face \, ijk} = \big( 2 - (2 + \nu) (1-\xi_i - \xi_j)^\nu \big) \xi_i \xi_j \big(\mathbf{e}_i - \mathbf{e}_j ) \\ 
    + \big( \xi_k^{\nu+1} - \xi_k \big) \big(\xi_i \mathbf{e}_i - \xi_j \mathbf{e}_j +  (\xi_j - \xi_i) \mathbf{e}_k \big),
    \label{eq:sing-edge-rot-curl-face-ijk}
\end{multline}
and similarly on face $ij\ell$ where $\xi_k = 0$,
\begin{multline}
    \nabla \times \mathbf{N}^{ijkl,r_1} \bigg |_\mathrm{face \, ij\ell} = \big( 2 - (2 + \nu) (1-\xi_i - \xi_j)^\nu \big) \xi_i \xi_j \big(\mathbf{e}_i - \mathbf{e}_j ) \\ 
    + \big( \xi_\ell^{\nu+1} - \xi_\ell \big) \big(\xi_i \mathbf{e}_i - \xi_j \mathbf{e}_j \big).
    \label{eq:sing-edge-rot-curl-face-ijl}
\end{multline}
In this case, we see that the first line of (\ref{eq:sing-edge-rot-curl-face-ijk}) and (\ref{eq:sing-edge-rot-curl-face-ijl}) provides the desired $\rho^\nu$ behavior up to the $\xi_i \xi_j$ weighting factor. The unit vector in the direction tangential to the edge is given in (\ref{eq:tan-dir}), and can be expressed in shorthand form as $\mathbf{\hat{u}}_{ij} = (\mathbf{e}_i - \mathbf{e}_j) / |\mathbf{e}_i - \mathbf{e}_j|$. Therefore, this term is parallel to edge $ij$, as is desired to match the behavior in (\ref{eq:sing-H-approx}).

As in (\ref{eq:sing-edge-grad-tan}), there is also a constant term with the $\xi_i \xi_j$ weighting factor which can be eliminated in combination with standard higher-order basis functions. Additionally, we have a $\xi_k^{\nu+1} - \xi_k$ term in (\ref{eq:sing-edge-rot-curl-face-ijk}) and a similar $\xi_\ell^{\nu+1} - \xi_\ell$ in (\ref{eq:sing-edge-rot-curl-face-ijl}). Similarly to the $\xi_k - \xi_k^\nu$ term in (\ref{eq:sing-edge-grad-expand-tan}), these terms are typically relatively small, and thus the behavior is dominated by the $\rho^\nu$ term parallel to the edge. Generalizing this approximation to the curl throughout the volume in (\ref{eq:sing-edge-rot-curl-full}) and taking the tangential part leads to the form provided in (\ref{eq:sing-edge-rot-curl-approx}) in Section \ref{subsubsec:edge-funcs}.




\bibliographystyle{IEEEtran}
\bibliography{citations}






\end{document}